\documentclass[reprint,nofootinbib,superscriptaddress, aps,longbibliography]{revtex4-1}
\usepackage{srcltx,graphicx}
\usepackage{amsmath, amssymb, amsthm}
\usepackage{color, xcolor, colortbl}
\usepackage{array, multirow}
\usepackage{hyperref}
\usepackage{epstopdf}
\usepackage{algorithm}
\usepackage{algorithmic}
\usepackage{caption}
\usepackage{subcaption}
\usepackage{color, colortbl}
\usepackage{braket}
\usepackage{hhline} 
\usepackage[normalem]{ulem}
\usepackage{placeins}
\RequirePackage[capitalize,nameinlink]{cleveref}[0.19]

\graphicspath{{images/}}

\floatname{algorithm}{Algorithm}

\newcommand\bbR{\mathbb{R}}
\newcommand\bbC{\mathbb{C}}

\DeclareMathOperator*{\argmin}{arg\,min} 

\definecolor{LightBlue}{rgb}{0.384,0.588,1}
\definecolor{LightGreen}{rgb}{0.443,1,0.384}
\definecolor{LightMagenta}{rgb}{0.984,0.435,1}
\definecolor{LightRed}{rgb}{1,0.435,0.435}

\usepackage[top=1in, bottom=1in, left=1in, right=1in]{geometry}

\newwrite\bibnotes
    \def\bibnotesext{Notes.bib}
    \immediate\openout\bibnotes=\jobname\bibnotesext
    \immediate\write\bibnotes{@CONTROL{REVTEX41Control}}
    \immediate\write\bibnotes{@CONTROL{%
    apsrev41Control,author="08",editor="1",pages="1",title="0",year="1"}}
     \if@filesw
     \immediate\write\@auxout{\string\citation{apsrev41Control}}%
    \fi

\begin{document}

\title{Efficient hybridization fitting for dynamical mean-field theory via semi-definite~relaxation}
\author{Carlos~Mejuto-Zaera}
\affiliation{Department of Chemistry, University of California, Berkeley, CA 94720}
\author{Leonardo~Zepeda-N\'u\~nez}
\affiliation{Department of Mathematics, University of California, Berkeley, CA 94720}
\author{Michael~Lindsey}
\affiliation{Department of Mathematics, University of California, Berkeley, CA 94720}
\author{Norm~Tubman}
\affiliation{Quantum Artificial Intelligence Lab. (QuAIL), Exploration Technology Directorate,
NASA Ames Research Center, Moffett Field, CA 94035, USA}
\author{Birgitta~Whaley}
\affiliation{Department of Chemistry, University of California, Berkeley, CA 94720}
\author{Lin~Lin}
\affiliation{Department of Mathematics, University of California, Berkeley, CA 94720}
\affiliation{Computational Research Division, Lawrence Berkeley National Laboratory, Berkeley, CA 94720}

\begin{abstract}
We introduce a nested optimization procedure using semi-definite relaxation for the fitting step in Hamiltonian-based cluster dynamical mean-field theory (DMFT) methodologies. We show that the proposed method is more efficient and flexible than state-of-the-art fitting schemes, which allows us to treat as large a number of bath sites as the impurity solver at hand allows. We characterize its robustness to initial conditions and symmetry constraints, thus providing conclusive evidence that in the presence of a large bath, our semi-definite relaxation approach can find the correct set of bath parameters without needing to include \emph{a priori} knowledge of the properties that are to be described. We believe this method will be of great use for Hamiltonian-based calculations, simplifying and improving one of the key steps in cluster dynamical mean-field theory calculations. 
\end{abstract}


\maketitle

\section{Introduction}

Understanding the behavior of materials from first principles has been one of the most important challenges in the physical sciences since the advent of quantum mechanics. This is particularly difficult for so-called strongly correlated systems, for which a mean-field treatment is insufficient. In the pursuit of understanding emergent phenomena in the thermodynamic limit of such systems, embedding methods have played a major role. Some of the most prominent examples of embedding techniques are dynamical mean-field theory (DMFT)~\cite{Kotliar1996,Maier2005,Kotliar2006}, density-matrix embedding theory (DMET)~\cite{Knizia2012}, and self-energy embedding theory (SEET)~\cite{Zgid2015} among many others. These methods have been widely successful in unveiling the low energy properties of model Hamiltonians and real materials alike~\cite{Chen2014, Zheng2016,LeBlanc2015,Zheng2017,Sakai2018,Paul2018,Markov2019,Walsh2019a,Walsh2019b}.

The embedding methods previously mentioned share the same framework; 
they substitute the computationally intractable original system with a simpler model system. This model system is usually composed of an interacting impurity, or \emph{cluster}, and a non-interacting \emph{bath}. 
The cluster commonly represents a subset of the original system, while its interplay with the rest of the original system is encoded by the bath.
The objective is to find a model system whose low energy properties coincide with or indicate the properties of the original system of interest.
Among these methods, Green's function based approaches, such as Hamiltonian-based DMFT~\cite{Caffarel1994,Zgid2012, Wolf2014, Wolf2015,Go2017,Mejuto2019}, are widely used. A common feature of this kind of embedding methods is the need to optimize, or fit, some function representing a parametrization of the bath. 

Although this fitting step is just as important as the rest for the success of the embedding calculation, it has received comparatively little attention in the method development research and existing literature. Particularly in Hamiltonian-based DMFT approaches the fitting step plays a crucial role: it can be understood as the steering wheel that guides the exploration of parameter space that has to culminate in the identification of the appropriate model system. The performance of the fit can influence the final model system (the fixed point of the Hamiltonian parameters) and whether it is representative of the thermodynamic limit of the system under consideration. Unfortunately, this fitting problem is highly non-convex in the sense that the optimization often
gets trapped in local minima, which strongly depend on the choice of the initial guess. Thus, typical approaches 
rely on running the optimization several (perhaps many) times, using different randomly generated initial guesses, and keeping the 
parameters that achieve the lowest cost~\cite{Go2015}. Previous studies trying to alleviate this issue have met with partial success~\cite{Zgid2011}. 

In this work, we propose a new efficient and empirically robust optimization algorithm that leverages semi-definite relaxation (SDR) for the fitting problem in Hamiltonian-based cluster DMFT approaches.
The algorithm relies on convex relaxation~\cite{Candes2009,CandesTao2010,VandenbergheBoyd1996}, coupled to a novel nested approach that combines state-of-the-art conic solvers~\cite{Becker2011} with standard gradient-based optimization algorithms. It provides for a fitting routine that we show to be (a) systematically improvable, (b) more efficient than commonly used optimization methods, and (c) empirically robust to initial conditions and symmetry constraints. One of the main advantages of this novel method is the emergence of symmetries in the bath parameters reflecting the geometry of the cluster, which in previous methods had to be imposed in the optimization~\cite{Koch2008,Liebsch2011,Foley2019}. These symmetries generalize to large baths and clusters, which we can handle by using a state-of-the-art impurity solver, the adaptive sampling configuration algorithm (ASCI)~\cite{Tubman2016,Mejuto2019}. As a result, it can handle large bath sizes and asymmetric clusters with an unprecedented accuracy.

The structure of the paper is as follows: In section~\ref{sec:methods} we describe in detail the methodologies employed and introduced: section~\ref{sec:methods_DMFT} contains a detailed description of the DMFT method and the role of fitting within the approach, while in section~\ref{sec:methods_SDR} we introduce and describe our SDR approach to bath fitting. Then section~\ref{sec:res} presents the numerical results of this paper: section~\ref{sec:res_fit} characterizes the properties of the SDR fit, and section~\ref{sec:res_DMFT} presents its performance in DMFT calculations explicitly. The converged bath parameters for the results presented in section~\ref{sec:res_DMFT} can be found in the Supplemental Material~\cite{supp}.

\section{Methods}
\label{sec:methods}

\subsection{Dynamical mean-field theory}
\label{sec:methods_DMFT}


DMFT and its cluster extensions form a family of numerical methods that treat strongly correlated many-body systems non-perturbatively~\cite{Kotliar1996, Kotliar2006, Maier2005}. The approach originates from the study of lattice Hamiltonian systems in infinite dimensions~\cite{Metzner1989,Mueller1989a,Mueller1989b,Kotliar1992}. The main idea is to map a fully interacting lattice to a (many-site) Anderson impurity model, composed of an interacting cluster that conserves all original interactions of the system and a non-interacting bath. Doing so, one substitutes an intractable interacting many-body system with a simpler one that can be studied with existing numerical techniques, e.g. quantum Monte Carlo~\cite{Gull2011} or exact diagonalization approaches (ED)~\cite{Caffarel1994}. For this mapping to be useful, one seeks to define the impurity model such that its low energy physics coincides with that of the original system. The precise way in which this matching is defined and achieved is briefly summarized later. First, we outline the key differences between the two main solver types for DMFT: Monte Carlo and Hamiltonian-based methods. The results in this work are relevant for the latter.

The DMFT algorithm differs significantly between Monte Carlo-based
solvers and Hamiltonian-based ones. The main advantage of using a Monte
Carlo solver is that one can work in the infinite-bath limit, as
required for formally exact implementation of DMFT. By contrast,
Hamiltonian-based methods need to truncate the bath, adding an extra
approximation to the calculation and raising questions about convergence
with respect to the bath size. However, Monte Carlo methods are formally
limited to computing quantities along the imaginary time/frequency axis,
thus they need to perform an analytical continuation to obtain spectral
quantities. Hamiltonian-based methods, meanwhile, can directly provide
results on both the imaginary and real axes. Moreover, while Monte Carlo
methods can handle the largest interacting clusters, they are limited by
the sign problem to relatively high temperatures and low doping rates. In summary, both families contribute to the calculation of phase diagrams. 

In this work, we address difficulties found in the finite-bath Hamiltonian approaches, where the bath truncation necessitates an optimization (fit) step. A brief description of the algorithmic structure follows. For a review of the DMFT algorithm using Hamiltonian-based methods, the reader is referred to~\cite{Zgid2011}.

In DMFT one first selects a subset of the original degrees of freedom, which are referred to as the cluster. The impurity model, defined below in Eq.~\eqref{eq:Himp}, includes these degrees of freedom and all their interactions, while neglecting the rest of the original system. To account for the coupling between the cluster and the rest of the lattice, the impurity model is completed with a set of non-interacting degrees of freedom, which we call {\it baths}, which are coupled to the cluster sites. A possible physical interpretation considers the baths as paths outside the cluster that a particle may traverse in the original system. Thus a possible motion of a particle starting inside the cluster, leaving it, moving through the rest of the lattice, and returning to the cluster may be represented in the impurity model as the particle hopping from a cluster site to a bath site and back. 

The Hamiltonian for the impurity model $H_{imp}$ is given by 
\begin{equation} \label{eq:Himp}
        H_{imp} = H_{C} + \sum_{\ell=1}^{N_b}\epsilon_\ell\  d^\dagger_{\ell}d_{\ell} + \sum_{\ell=1}^{N_b}\sum_{\alpha=1}^{N_c}\left(V_{\alpha,\ell}\ d^\dagger_{\ell}c_{\alpha} + \mathrm{h.c.}\right)\ ,
\end{equation}
where $N_c$ is the number of cluster degrees of freedom, $N_b$ the number of baths, $H_C$ is the original Hamiltonian restricted to the cluster degrees of freedom; $c_\alpha$, and $d_\ell$ correspond to the cluster and bath annihilation operators respectively. The bath degrees of freedom are characterized by their single particle energies $\epsilon_\ell$ and their couplings $V_{\alpha,\ell}$ to the cluster sites. The goal of DMFT is to determine the bath parameters $\{\epsilon_\ell\}_{\ell=1}^{N_{b}}$ and $\{V_{\alpha,\ell}\}_{\ell=1, \alpha =1}^{N_b, N_c}$ such that the low energy physics of $H_{imp}$ and the original system $H$ coincide. This means that the Green's function $G_{imp}(\omega)_{\alpha, \beta}$ of $H_{imp}$ should be the same as the corresponding submatrix of the Green's function of the global Hamiltonian~$H$. This is achieved in DMFT by a self-consistent loop. 
To this effect, Hamiltonian-based methods, e.g., ED~\cite{Caffarel1994} or selective configuration interaction solvers~\cite{Zgid2012, Go2017, Mejuto2019}, concentrate on the hybridization function $\Delta^{Bath}$. This is the part of the non-interacting Green's function generated by the bath degrees of freedom. It can be shown~\cite{Phillips2012} that the hybridization function follows the analytical expression

\begin{equation} \label{eq:Hyb}
    \Delta^{Bath}(\omega)_{\alpha,\beta} = \sum_{\ell=1}^{N_b}\frac{V_{\alpha,\ell} V_{\beta,\ell}^*}{\omega - \epsilon_\ell} .
\end{equation}

Note that $\Delta^{Bath}$ equivalently encodes the fitting parameters $\{\epsilon_\ell\}_{\ell=1}^{N_{b}}$ and $\{V_{\alpha,\ell}\}_{\ell=1, \alpha =1}^{N_b, N_c}$ as a single rational function. The self-consistent loop is defined by matching $G_{imp}$, the impurity model's Green's function, to $G_{latt}$, the Green's function of the original system. Hence to close the loop we need a procedure to estimate $G_{latt}$ from $G_{imp}$, which is detailed in the next paragraph. From $G_{latt}$ we then extract the hybridization function $\Delta^{Calc}$ which should match $\Delta^{Bath}$ at self-consistency. 

 A given set of $\{\epsilon_\ell\}$ and $\{V_{\alpha,\ell}\}$ defines the impurity Hamiltonian $H_{imp}$, so that one can compute $G_{imp}$, at zero temperature, as follows,

\begin{equation} \label{eq:GF}
  \begin{split}
  G_{imp}(\omega)_{\alpha,\beta} =& \bra{\psi_0}c_{\alpha}\frac{1}{\omega - (H_{imp} - E_0) + i\eta}c^\dagger_\beta\ket{\psi_0} \\ 
  &+ \bra{\psi_0}c^\dagger_{\beta}\frac{1}{\omega + (H_{imp} - E_0) - i\eta}c_\alpha\ket{\psi_0},
  \end{split}
\end{equation}
where $E_0$ is the ground state energy, $\ket{\psi_0}$ is the ground state wave function, and $\eta$ is an infinitesimal positive broadening factor. From this, one can compute the cluster self energy,

\begin{equation} \label{eq:SE}
   \begin{split}
    \Sigma_c(\omega)_{\alpha, \beta} =& \ (\omega + \mu + i\eta)\delta_{\alpha,\beta} - h_{C,\alpha,\beta} \\&- G_{imp}^{-1}(\omega)_{\alpha,\beta} - \Delta^{Bath}(\omega)_{\alpha,\beta},
   \end{split}
\end{equation}
where $h_{C}$ is the non-interacting part of $H_C$ and $\mu$ is the chemical potential. The DMFT approximation amounts to assuming that one can estimate the full lattice Green's function by using the \emph{local} cluster self energy, according to
\begin{equation} \label{eq:GFlatt}
    G_{latt}(\mathbf{k}, \omega) = \left[(\omega + \mu + i\eta)-h(\mathbf{k})-\Sigma_c(\omega)\right]^{-1},
\end{equation}
where $h(\mathbf{k})$ is the Fourier transform of $h_{C}$ into momentum space. This recipe for computing $G_{latt}$ corresponds to a block-diagonal ansatz for the global self-energy in real space, with blocks corresponding to the clusters. Eq. \eqref{eq:GFlatt} is exact in the infinite-dimensional limit for the Hubbard model~\cite{Kotliar1996}, even when the cluster is only comprised of one single site. In finite dimensions, there are further corrections to Eq.~\eqref{eq:GFlatt}; these terms incorporate the non-locality of the self-energy beyond the cluster. To improve the DMFT approximation one can consider larger and larger clusters, the exact result being recovered in the limit of infinite cluster size. The finite size scaling of the error is $\mathcal{O}\left(\frac{1}{N_c}\right)$ for cluster DMFT~\cite{Maier2005}. Other cluster methods have better scaling, such as $\mathcal{O}\left(\frac{1}{N_c^2}\right)$ for the dynamical cluster approximation (DCA), at the cost of more complex bath structures~\cite{Koch2008}. In particular, it can be shown that in cluster DMFT the only cluster sites that have non-zero couplings to the bath sites are those on the boundary of the cluster. This reduces the number of fit parameters and simplifies the structure of the Hamiltonian in Eq.~\eqref{eq:Himp}. One can then Fourier-transform Eq.~\eqref{eq:GFlatt} to compute the original Green's function for the cluster degrees of freedom with the following expression,

\begin{equation} \label{eq:GFloc}
    G(\mathbf{R}_0, i\omega) = \frac{1}{V_{BZ}}\int_{BZ}\mathrm{d}\mathbf{k}\ \left[(i\omega + \mu)-h(\mathbf{k})-\Sigma_c(i\omega)\right]^{-1},
\end{equation}
where $BZ$ stands for the first Brillouin zone and $\mathbf{R}_0$ is the center of the cluster and the origin of the superlattice of clusters. The hybridization function can be extracted by solving for $\Delta$ in Eq.~\eqref{eq:SE} as 

\begin{equation} \label{eq:HybCalc}
   \begin{split}
   \Delta^{Calc}(\omega)_{\alpha,\beta} =
   & (\omega + \mu +i\eta)\delta_{\alpha,\beta} - h_{C,\alpha,\beta} \\ &- G^{-1}(\mathbf{R}_0, i\omega)_{\alpha,\beta} - \Sigma_c(\omega)_{\alpha, \beta} .
   \end{split}
\end{equation}
If $\Delta^{Calc}$ coincides with $\Delta^{Bath}$ in Eq. \eqref{eq:Hyb}, then the self-consistency loop terminates. Otherwise, one finds a new set of bath parameters $\{\epsilon_\ell\}$ and $\{V_{\alpha,\ell}\}$ by fitting Eq. \eqref{eq:Hyb} to $\Delta^{Calc}$ and starts a new iteration with a new impurity Hamiltonian. The cost function to minimize is usually defined in terms of a matrix norm of the difference between $\Delta^{Bath}$ and $\Delta^{Calc}$, evaluated on some frequency grid $\{ i\omega_n\}$ on the imaginary frequency axis. One avoids thus the poles of the Green's function and makes the fitting problem much easier. For simplicity, in this paper we will always consider the Frobenius norm, denoted $\Vert \,\cdot\, \Vert_F$, and a cost function of the form

\begin{equation} \label{eq:CostFunc}
   \begin{split}
    \mathcal{J}(\{\epsilon_\ell\},\{V_{\alpha,\ell}\}&) = \\& \frac{1}{N_\omega}\sqrt{\sum_{n=1}^{N_\omega} \left\| \Delta^{Calc}(i\omega_n) - \Delta^{Bath}(i\omega_n)\right\|_F^2} .
   \end{split}
\end{equation}

 We point out that it is possible to add a frequency dependent weight function $g(\omega)$ to the sum. The objective of this weight function is to preferentially fit the frequencies close to the real axis. In this work, however, we emphasize the imaginary low-frequency region by choosing frequency grids $\left\{\omega_n\right\}_{n=1}^{N_\omega}$ with different point densities. Thus, we keep  $g(\omega) = 1$ throughout the discussion. We will show that once enough bath parameters are provided, this fine tuning has no qualitative impact on our proposed optimization method.
 
 Even when using the simplified cost function in Eq.~\eqref{eq:CostFunc}, the optimization problem at hand is highly nontrivial and has been object of diverse studies~\cite{Koch2008,Senechal2010,Liebsch2011}. We are tasked with minimizing the difference between two complex-valued, frequency-dependent $N_c\times N_c$ matrices using $N_b\cdot(N_c + 1)$ parameters. To maximize the model accuracy, $N_c$ and $N_b$ should be taken as large as our impurity solver allows. But the larger $N_c$ and $N_b$ are, the more difficult the fitting problem becomes, and in particular standard optimization procedures becomes increasingly susceptible to falling into local minima. In fact, when using a very efficient impurity solver, i.e., a solver capable of accommodating fairly large values of $N_c$ and $N_b$, an inefficient fitting step can become the computational bottleneck~\cite{Mejuto2019}.  In state-of-the-art Hamiltonian-based DMFT codes it is the norm to use elaborate fitting schemes in order to avoid instabilities and local minima~\cite{Zgid2011,Go2015}. Furthermore, due to fitting on the imaginary axis, the result of the DMFT self-consistency can be significantly dependent on the fit frequency range~\cite{Mejuto2019}, since physical information is distributed unevenly there (i.e. sum rules being encoded in the high frequency limit while the details of the state properties being encoded close to the real axis). Additionally, traditional approaches tend to require clusters of high symmetry in order to reduce the complexity of the fitting problem by block-diagonalizing the Green's function~\cite{Koch2008}, which limits the size and shape of the clusters  that can be studied.

In the following sections, we present a novel fitting method for impurity problems based on semi-definite relaxation (SDR) that addresses the issues mentioned in the previous section and substantially improves upon existing methodologies. Our method~(i) implements the most general structure for a finite bath hybridization~\cite{Bolech2003},~(ii) shows fast timings allowing for treatment of large $N_c$ and $N_b$,~(iii) is systematically improvable in $N_b$,~(iv) is empirically robust to starting conditions and symmetry constraints, and~(v) shows rapid convergence for the low energy baths $\epsilon_\ell$. We demonstrate our claims by presenting results from zero-temperature cluster DMFT calculations using our fitting method and the  adaptive sampling configuration interaction (ASCI) algorithm as impurity solver~\cite{Tubman2016,Tubman2018,Tubman2018a, Mejuto2019}

\subsection{Formulation of the hybridization fitting}
\label{sec:methods_SDR}
Motivated by the previous section, let ${\Delta:\bbC \rightarrow \bbC^{N_{c}\times N_{c}}}$ be a function that we seek to approximate in a discrete subset of the imaginary axis that we denote by $\Omega$. In particular, we aim to find an approximation of the form 
\begin{equation} \label{eq:ansatz}
        \Delta(i\omega) \approx \sum_{\ell =1}^{N_{p}} \frac{V_\ell V_\ell^*}{i\omega - \lambda_\ell},
\end{equation}
where $\omega \in \bbR$. We have introduced a shorthand notation omitting the cluster indices $\{\alpha,\beta\}$, so that the hybridization matrix is written simply as $\Delta$, the coupling amplitudes are organized in vectors $V_\ell \in \bbC^{N_c}$, and the bath energies are still treated as scalars $\lambda_\ell \in \bbR$  for $\ell = 1, \ldots, N_{p}$. In addition, for reasons to be clarified later, we denote the poles in Eq.~\eqref{eq:ansatz} by $\lambda_\ell$, not by $\epsilon_\ell$ as in the last section. 

To find the approximation in Eq.~\eqref{eq:ansatz} we define the cost function, 
\begin{equation} \label{eq:cost}
        \mathcal{J}(\{ \lambda_\ell, V_\ell \}_{\ell=1}^{N_{p}}) =  \frac{1}{N_\omega}\sqrt{  \sum_{n = 1}^{N_{\omega}} \left  \| \Delta(i\omega_n)  - \sum_{\ell=1}^{N_{p}} \frac{V_\ell V_\ell^*}{i\omega_n - \lambda_\ell}   \right \|_{F}^2 },
\end{equation}
where $\Omega = \{i\omega_n\}_{n = 1}^{N_{\omega}}$. We aim to solve the optimization problem
\begin{equation} \label{eq:optimization}
         \min_{\lambda_\ell, V_\ell} \mathcal{J}(\{\lambda_\ell, V_\ell \}_{\ell=1}^{N_{p}}),
\end{equation}
whose minimizer provides the parameters for the ansatz in Eq.~\eqref{eq:ansatz}. 

One natural approach to solve Eq.~\eqref{eq:optimization} is to use gradient based optimization  to find the parameters, for a fixed number of poles, $N_{p}$. As already stated, the problem is highly non-convex in the sense that the optimization often
gets trapped in local minima, which strongly depend on the choice of the initial guess. If one were to fix $\{ V_\ell \}_{\ell=1}^{N_{p}}$, then the resulting problem could be reduced to a scalar-valued rational approximation problem that can be solved efficiently~\cite{Damle2013}. However, such algorithms, which are designed for scalar functions, fail to be directly applicable to our matrix-valued setting. 

The main observation for solving Eq.~\eqref{eq:optimization} is that if we
fix the poles $\{\lambda_\ell \}_{\ell=1}^{N_{p}}$, then the resulting problem
can be reformulated using convex relaxation
\cite{Candes2009,CandesTao2010,VandenbergheBoyd1996} into an SDR problem. The
reformulated SDR problem is convex and can be efficiently solved using
conic solvers \cite{Becker2011}, relying either on interior-point
methods \cite{Sturm1999} or first-order methods \cite{ODonoghue2016}. By
exploiting this observation, we develop an algorithm with the following steps.

\begin{enumerate}
\item Relax the ansatz in Eq.~\eqref{eq:ansatz} by replacing the rank-one matrices $V_\ell V_\ell^*$ by positive semi-definite matrices $X_\ell$ of arbitrary rank, resulting in a new loss function,

\item Optimize the poles and the matrices in an alternating fashion, which leverages state-of-the-art conic programming and quasi-Newton methods, and 

\item Truncate the resulting positive semi-definite matrices and expanding them into sums of rank-one matrices, obtaining an approximation satisfying the ansatz in Eq.~\eqref{eq:ansatz}.

\end{enumerate}

Finally, it should be noted that in cluster DMFT, only the $N^{'}_c$ sites in the boundary of the cluster have non-vanishing $V_{\ell, \alpha}$~\cite{Koch2008}. In these cases, we can restrict the fit to a function $\Delta:\bbC \rightarrow \bbC^{N^{'}_{c}\times N^{'}_{c}}$.

\subsubsection{Relaxation}

We observe that the outer product $V_\ell V_\ell^*$ defines a positive semi-definite matrix of rank one. In a nutshell, the relaxation consists of replacing the rank-one matrices by more general positive semi-definite matrices $X_\ell$ without rank restriction, yielding the more general ansatz 

\begin{equation} \label{eq:SDR_ansatz}
        \Delta(i\omega) \approx \sum_{\ell =1}^{N_{p}} \frac{X_\ell}{i\omega - \lambda_\ell}.
\end{equation}

Analogous to the loss function defined in Eq.~\eqref{eq:cost}, we define the following loss function:
\begin{equation} \label{eq:spd_loss}
   \begin{split}
        \mathcal{J}_{SDR}(\{ \lambda_\ell, X_\ell\}&_{\ell=1}^{N_{p}}) =\\& \frac{1}{N_\omega}\sqrt{  \sum_{n = 1}^{N_{\omega}} \left \| \Delta(i\omega_n) - \sum_{\ell =1}^{N_{p}} \frac{X_\ell}{i\omega_n - \lambda_\ell} \right \|_F ^2}.
   \end{split}
\end{equation}
We replace the optimization problem in Eq.~\eqref{eq:optimization} by a relaxed optimization problem
\begin{equation} \label{eq:optimization_spd}
         \min_{\lambda_\ell, X_\ell \succeq 0 } \mathcal{J}_{SDR}(\{ \lambda_\ell, X_\ell\}_{\ell=1}^{N_{p}}), 
\end{equation}
where $ X_\ell \succeq 0$ means that $X_\ell$ is a Hermitian positive semi-definite matrix. This constraint is usually called a \emph{conic} constraint, due to the fact that the set of Hermitian positive semi-definite matrices forms a convex cone \cite{VandenbergheBoyd1996}, i.e., if $A \succeq 0 $ and  $B\succeq 0$, then $a A + b B \succeq 0$, for any non-negative real numbers $a$ and $b$.

From the standpoint of the size of the optimization space, we have increased the number of unknowns from $ \mathcal{O}(N_{p} N_{c})$ to $ \mathcal{O}( N_{p} N_{c}^2)$. In addition, due to the rank relaxation, the minimizer from Eq.~\eqref{eq:optimization_spd} does not immediately yield a function of the form given by Eq.~\eqref{eq:ansatz}. However, this shortcoming is not fundamental and can be easily addressed as follows at minimal additional cost.

Suppose that $\{ \hat{\lambda}_\ell, \hat{X}_\ell\}_{\ell=1}^{N_{p}}$ are the minimizers of Eq.~\eqref{eq:optimization_spd}. By construction we have that $\hat{X}_\ell \succeq 0 $. Then there exist matrices $\{U_\ell\}_{\ell=1}^{N_{p}}$ such that
\begin{equation}
        \hat{X}_\ell = U_\ell U^*_\ell = \sum_{q = 1}^{N_{c}}  U_{\ell,q} U^*_{\ell, q},
\end{equation}
where $U_{\ell,q}$ is the $q$-th column of $U_\ell$. 
Then we can expand 
\begin{equation} \label{eq:effective_baths}
        \sum_{\ell=1}^{N_{p}} \frac{\hat{X}_\ell}{i\omega - \hat{\lambda}_\ell} = \sum_{\ell=1}^{N_{p}} \sum_{q=1}^{N_{c}}  \frac{U_{\ell,q} U^*_{\ell,q}}{i\omega - \hat{\lambda}_\ell} = \sum_{r=1}^{N_{b}^{\text{eff}}} \frac{ \tilde{V}_r \tilde{V}^{*}_r}{i\omega - \tilde{\epsilon}_r}, 
\end{equation}
where in the last expression we have adopted appropriate definitions for $\tilde{V}_r$, $\tilde{\epsilon}_r$, and the index $r$. It is clear that Eq.~\eqref{eq:effective_baths} has the same form as the ansatz in Eq.~\eqref{eq:ansatz}, though here the \emph{effective} number of baths, $N_{b}^{\text{eff}}$, is equal to $N_{c} N_{p}$. In particular, each bath energy $\tilde{\epsilon}_r$, is associated to $N_{c}$ different baths.

\subsubsection{Optimization}

The minimization in Eq.~\eqref{eq:optimization_spd} is non-trivial due mainly to the conic constraint. Unfortunately, popular methods such as the alternating direction method of multipliers (ADMM)~\cite{Boyd2011}, the method of alternating direction (AMA)~\cite{Tseng91}, or backward-forward splitting (BFS) schemes~\cite{Combettes2011} are not readily applicable to our case, given the non-convex nature of the cost function. Thus we propose a nested algorithm as follows.

Consider a loss function with respect to the poles as
\begin{equation} \label{eq:poles_cost}
        \mathcal{J}_{\text{pol}}(\{ \lambda_\ell\}_{l=1}^{N_{p}}) = \min_{X_\ell \succeq 0} \mathcal{J}_{\text{SDR}}(\{ \lambda_\ell, X_\ell\}_{\ell=1}^{N_{p}}),
\end{equation}
obtained by fixing the poles and solving an SDR problem using conic solvers \cite{mosek2005,SDPT32003}.
Using the optimality conditions for this problem the gradient can be easily computed via 
\begin{equation} \label{eq:nabla_L}
  \begin{split}
  \frac{\partial}{\partial \lambda_k} \mathcal{J}_{\text{pol}}(&\{ \lambda_\ell\}_{\ell=1}^{N_{p}}) =\\&  \frac{\partial}{\partial \lambda_k} \left . \left (  \sqrt{  \sum_{j = 1}^{N_{\omega}} \left  \| \Delta(i\omega_j) - \sum_{\ell =1}^{N_{p}} \frac{X_\ell}{i\omega_j - \lambda_\ell}    \right \|_F^2} \right ) \right |_{X_\ell = X'_\ell},
  \end{split}
\end{equation}
where $X'_\ell = \argmin_{X_\ell \succeq 0} \mathcal{J}_{\text{SDR}}(\{ \lambda_\ell, X_\ell\}_{\ell=1}^{N_{p}})$.

Using this newly defined cost function we solve
\begin{equation} \label{eq:min_lambda}
  \{\hat{\lambda}_\ell\}_{\ell=1}^{N_{p}}=\argmin_{\{\lambda_\ell\}_{\ell=1}^{N_{p}} \subset \bbR } \mathcal{J}_{\text{pol}}(\{ \lambda_\ell\}_{\ell=1}^{N_{p}})
\end{equation}
using standard unconstrained gradient based methods. The full optimization loop is summarized in Alg.~\ref{alg:nested_minimization}, which was implemented in MATLAB, where CVX~\cite{cvx} is used to solve Eq.~\eqref{eq:poles_cost}, the derivatives of $\mathcal{J}_{\text{pol}}$ are computed analytically using Eq.~\eqref{eq:nabla_L}, and the outer optimization loop is solved with the BFGS method \cite{Broyden1970} included in the \texttt{fminunc} routine. We point out that the present algorithm is loosely related to BFS schemes \cite{Combettes2011,LionsMercier1973}, and that it can be considered as an inexact alternating descent algorithm~\cite{Ortega2000}.

\begin{algorithm}[H]
\begin{algorithmic}
\REQUIRE Initial guess $\{\lambda_\ell\}_{\ell =1}^{N_{p}}$
\ENSURE $ \{ \hat{\lambda}_\ell, \hat{X}_\ell\}_{\ell =1}^{N_{p}}$
 \WHILE{$\Vert \nabla \mathcal{J}_{\text{pol}}(\{\lambda_\ell\}_{\ell=1}^{N_{p}})\Vert > \epsilon_{\text{tol}}$}
 \STATE $X'_\ell \leftarrow \argmin_{X_\ell \succeq 0} \mathcal{J}_{\text{SDR}}(\{ \hat{\lambda}_\ell, X_\ell\}_{\ell=1}^{N_{p}})$ by solving the SDR problem in Eq.~\eqref{eq:poles_cost}
 \STATE $\lambda_\ell \leftarrow \text{BFGS}(\lambda_\ell, \nabla
 \mathcal{J}_{\text{pol}}(\{ \lambda_\ell\}_{\ell=1}^{N_{p}}))$, where $\nabla
 \mathcal{J}_{\text{pol}}$ is computed using
 Eq.~\eqref{eq:nabla_L} and $X'_\ell$
 \ENDWHILE
 \STATE $\hat{\lambda}_\ell \leftarrow \lambda_\ell$
 \STATE $\hat{X}_\ell \leftarrow \argmin_{X_\ell \succeq 0} \mathcal{J}_{\text{SDR}}(\{ \hat{\lambda}_\ell, X_\ell\}_{\ell=1}^{N_{p}})$
 \end{algorithmic}
 \caption{Pseudo-code for the nested optimization routine.}
 \label{alg:nested_minimization}
\end{algorithm}

\subsubsection{Truncation}

As explained above, it is possible to post-process the output of Alg.~\ref{alg:nested_minimization} in order to obtain an approximation satisfying the ansatz in Eq.~\eqref{eq:ansatz}. However, this procedure increases the effective bath size significantly, which in turn increases the computational cost of the DMFT loop by enlarging the impurity problems. Fortunately, in many cases, the contributions of many of these effective baths are extremely small and can be discarded.

Such a truncation can be performed efficiently by solving one eigenvalue problem per SDR matrix. We compute the eigenvalue decomposition of $\hat{X}_\ell$ as
\begin{equation}
        \hat{X}_\ell = Q_\ell \Lambda_\ell Q^*_\ell
\end{equation}
where $Q_\ell$ are unitary matrices and $\Lambda_\ell$ is a real diagonal matrix, containing the eigenvalues of $X_\ell$ ordered non-increasingly. For a fixed threshold $\delta$, which we set to $10^{-5}$ in this work, we gather all the eigenvalues above this threshold and their corresponding eigenvectors in the matrices $\Lambda^{\delta}_\ell \in \bbR^{N^{\delta}_{\ell}\times  N^{\delta}_{\ell}}$ and
$Q_\ell^{\delta} \in \bbC^{N_{c} \times  N^{\delta}_{\ell}}$ , respectively, where $N^{\delta}_{\ell}$ is the number of eigenvalues of $\hat{X}_\ell$ greater than $\delta$, namely the `$\delta$-rank' of $\hat{X}_\ell$.
Then we define 
\begin{equation}
        U^{\delta}_\ell = Q_\ell^{\delta} (\Lambda^{\delta}_\ell)^{1/2}, \text{ such that} \quad \| U^{\delta}_\ell (U^{\delta}_\ell)^{*} - \hat{X}_\ell \|_F \lesssim \delta 
\end{equation}
Finally, we can follow Eq.~\eqref{eq:effective_baths} to obtain an approximation satisfying the ansatz in Eq.~\eqref{eq:ansatz}, but, hopefully, with somewhat smaller effective bath size, which is equal to the sum of the $\delta$-ranks of the semi-definite matrices $\hat{X}_\ell$. We point out that it should be possible to obtain a smaller effective bath size, by adding a rank-decreasing penalty, such as the nuclear norm, or trace, to the definition of $\mathcal{J}_{\text{pol}}$. However, we were not able to reliably achieve this without significantly decreasing the fit accuracy.  
\textbf{}
To conclude, we remark that we have not explicitly said how we initialize the poles in Alg.~\ref{alg:nested_minimization}. This will be stated in section~\ref{sec:res}.

\subsection{Impurity solver - ASCI}

Before turning to the analysis of the results, a few words on the impurity solver are in order. As mentioned previously, to obtain accurate results in cluster DMFT it is important to raise $N_c$ and $N_b$. In turn, this demands an impurity solver able to compute the Green's function $G_{imp}$ in Eq.~\eqref{eq:GF} for a system with as many strongly-correlated degrees of freedom as possible. For zero temperature, Hamiltonian-based DMFT calculations, exact diagonalization (ED)~\cite{Caffarel1994} and truncation schemes \cite{Zgid2012,Go2017,Mejuto2019} have been very popular~\cite{Go2015,Wang2017,Sakai2018}. The main idea behind truncation methods is the projection of the full system Hamiltonian into a subset $\mathcal{T}$ of the complete Hilbert space $\mathcal{H}$. This subspace is chosen as small as possible, while still being able to represent the full ground state wave function accurately. 

Among this kind of truncated ED schemes, the adaptive sampling configuration interaction (ASCI) algorithm~\cite{Tubman2016,Tubman2018,Tubman2018a} has been shown recently to be able to treat impurity models with large $N_c$ and $N_b$ efficiently~\cite{Mejuto2019}. The key aspect of ASCI is the identification of the optimal Hilbert space truncation of a given size to represent the ground state wavefunction of a many-body system. This is done by introducing a ranking that allows one to estimate the most relevant Hilbert space degrees of freedom for the description of the ground state.
ASCI is expected to perform particularly well in systems with (a) a high Hamiltonian-connectivity and (b) a ground state mainly composed by a few states in some single-body basis. In this work, using ASCI as impurity solver allows to use $N_p$ up to 6 in the SDR fit, making the study of the large bath behavior of the fit possible, which is the regime where SDR becomes the more useful.

\begin{table*}[t!]
  \begin{center}
    \begin{tabular}{|l|p{14cm}|} 
    \hline
    \textbf{Symbol} & \textbf{Meaning} \\
    \hline\hline
    $N_c$       & Number of cluster sites in the cluster DMFT calculation.\\
    \hline
    $N^{'}_c$   & Number of sites in the cluster boundary. Only this have nonzero coupling elements $V_{\alpha,k}$~\cite{Koch2008}.\\
    \hline
    $N_p$       & Number of poles in the SDR fit. \\
    \hline
    $N_b$       & Number of bath sites in the cluster DMFT calculation. \\
    \hline
    $N^{\text{eff}}_b$ & Number of effective bath sites in the SDR fit. This number is determined after the truncation step, its maximal value being $N^{'}_c\cdot N_p$. \\
    \hline
    \end{tabular}
  \end{center}
  \caption{Overview of notation.}
  \label{Table:notation}
\end{table*}

\section{Results}
\label{sec:res}
\subsection{Characterization of the fit}
\label{sec:res_fit}


In this section, we study in detail the performance of the SDR fit in Alg.~\ref{alg:nested_minimization} for impurity models, outside of the full DMFT calculation. We use as input data sample hybridization function matrices $\Delta$ from cluster DMFT calculations in (2$\times$2), (3$\times$3) and (4$\times$4) clusters of the two dimensional square lattice one-band Hubbard model. The Hamiltonian of this system is given by
\begin{equation} \label{eq:Hubbard}
        H = -t\sum_{\langle \alpha,\beta \rangle, \sigma} c^\dagger_{\alpha,\sigma} c_{\beta, \sigma} -\mu \sum_\alpha n_\alpha + U \sum_{\alpha} n_{\alpha, \uparrow} n_{\alpha, \downarrow},
\end{equation}
where $t$ is the hopping amplitude, $\langle \alpha, \beta \rangle$ denotes nearest neighbours, $\sigma = \{\uparrow,\downarrow \}$ is the spin degree of freedom, $\mu$ the chemical potential, $U$ the local Coulomb interaction, $n_{\alpha,\sigma}$ is the number counting operator for site $\alpha$, and spin species $\sigma$ and $n_\alpha = n_{\alpha, \uparrow} + n_{\alpha, \downarrow}$. Unless otherwise specified, $U\ / \  t = 8$ throughout the calculations.
We aim to showcase the following properties: 
\begin{enumerate}
    \item The quality of the SDR fit, as measured by the error cost function, can be systematically improved by increasing the number of poles.
    \item The SDR fitting method is significantly faster than other non-linear fitting procedures when applied to this kind of impurity models.
    \item The pole energies obtained from the SDR fit are robust with respect to the initial conditions. Furthermore, there is an easy diagnostic for identifying poor initial conditions, namely the rank of the $X_\ell$ matrices corresponding to each pole.
    \item The fitting results can be further improved by taking the SDR solution in the form given by Eq.~\eqref{eq:effective_baths} and using it as an initial guess for the optimization in Eq.~\eqref{eq:optimization}. This last optimization is implemented via gradient-based methods.
    \item The pole energies close to zero converge fairly rapidly with the number of poles.
\end{enumerate}

The first four points are to be desired from a numerical point of view, since they identify the SDR method as efficient and robust. The last point is of particular importance from a physical perspective, since it helps justify the bath truncation in Hamiltonian-based approaches to DMFT. Indeed, if the low energy baths converge rapidly with the number of poles, this means that the low energy behavior of our embedding Hamiltonian also converges rapidly with the number of baths. Eventually, adding extra bath degrees of freedom will exclusively affect the high energy features of the model, which may not be the main interest of a low temperature (zero temperature) study. This allows us to justify \emph{a posteriori} the bath truncation. Moreover, using SDR allows us in principle to determine the minimal number of bath orbitals required to converge the spectral properties of the embedding model up to the scale of interest. Whether the resulting bath size is then amenable to be treated with available impurity solvers, like the ASCI algorithm, is then something to be considered on a case by case basis.

The fits performed in this subsection are done over 50 imaginary frequency points in a linear span between $i\omega\ /\ t = 0$ and $i\omega\ /\ t = 40$. A detailed study of the effect that the frequency grid has on the fit is deferred to section~\ref{sec:res_DMFT}. Since we have introduced several notation elements, we summarize the main abbreviations in Tab.~\ref{Table:notation}.

\subsubsection{Fitting error}

\begin{figure*}[t]
\begin{subfigure}{0.5\textwidth}
  \includegraphics[width=\linewidth]{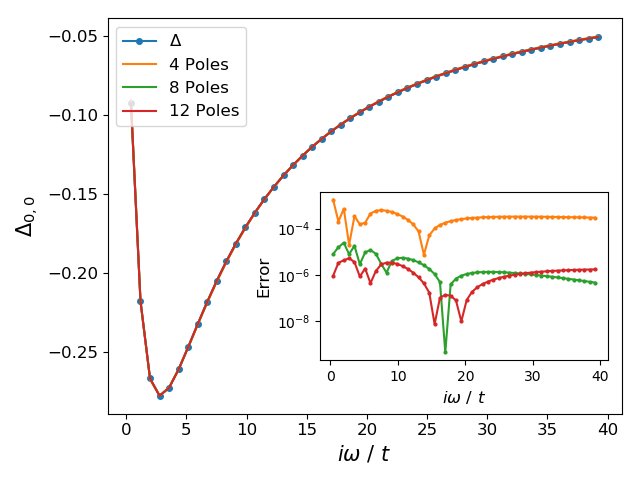}
  \caption{Diagonal component of $\Delta$.}
  \label{fig:diag_hyb}
\end{subfigure}%
\begin{subfigure}{0.5\textwidth}
  \includegraphics[width=\linewidth]{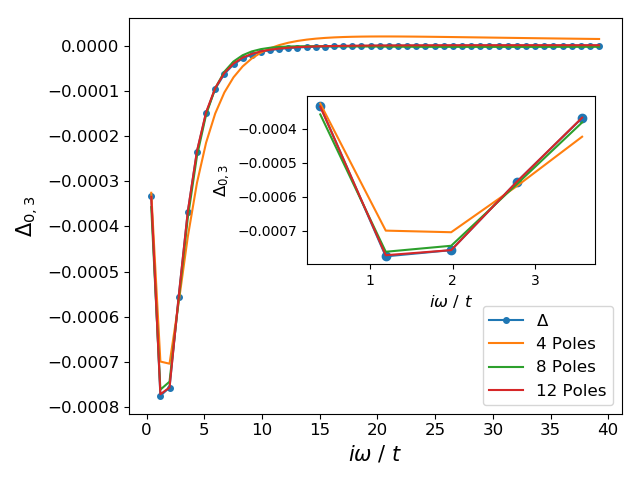}
  \caption{Off-diagonal component of $\Delta$.}
  \label{fig:offdiag_hyb}
\end{subfigure}
\caption{Fit results for the hybridization function $\Delta$ of a (2$\times$2) cluster impurity model for different number of poles. Shown are the imaginary parts of the $(0,0)$ diagonal and the $(0,3)$ off-diagonal components of the hybridization function to be fit (blue) and the resulting fits using 4 poles (orange), 8 poles (green) and 12 poles (red). The inset in the left subplot shows the absolute difference between the three fits and the computed hybridization function.}
\label{fig:fit_improv}
\end{figure*}

First, we want to examine how the fitting error progresses as a function of the number of poles for the different hybridization functions. We report the error and effective bath sizes in Tab. \ref{Table:errors}.

\begin{table}[h!]
  \begin{center}
    \begin{tabular}{|r||c|c||c|c||c|c|} 
    \hline
     & \multicolumn{2}{c||}{\textbf{(2$\times$2)}} &\multicolumn{2}{c||}{\textbf{(3$\times$3)}} & \multicolumn{2}{c|}{\textbf{(4$\times$4)}} \\
     \hline
      \textbf{$N_p$} & \textbf{$N^{\text{eff}}_b$} & \textbf{Error} & \textbf{$N^{\text{eff}}_b$} & \textbf{Error} & \textbf{$N^{\text{eff}}_b$} & \textbf{Error}\\
      \hline
         2  &   8   &    2.18e-3  & 16 & 2.16e-3 & 24 & 1.56e-2\\   
                 3  &   12  &    1.59e-3  & 24 & 1.57e-3 & 36 & 8.44e-3\\   
                 4  &   16  &    2.88e-4  & 32 & 3.97e-4 & 47 & 2.09e-3\\   
                 5  &   20  &    2.64e-4  & 40 & 2.93e-4 & 56 & 1.31e-3\\   
                 6  &   24  &    1.75e-5  & 48 & 5.99e-5 & 63 & 9.55e-4\\   
                 7  &   28  &    1.14e-5  & 52 & 4.04e-5 & 72 & 8.37e-4\\   
                 8  &   32  &    2.50e-6  & 55 & 5.41e-5 & 80 & 7.20e-4\\   
                 9  &   35  &    1.56e-6  & 62 & 5.26e-5 & 86 & 7.16e-4\\   
                10  &   38  &    1.50e-6  & 66 & 3.08e-5 & 88 & 7.07e-4\\   
                \hline
    \end{tabular}
  \end{center}
  \caption{Error and effective bath sizes $N^{\text{eff}}_b$ for hybridization function fits on three different cluster sizes  a function of the number of poles $N_p$ using the SDR fit.}
  \label{Table:errors}
\end{table}

From Tab. \ref{Table:errors} it becomes clear that adding poles systematically improves the fit. The dominant contribution to the imaginary part of all hybridization functions considered here is the diagonal, whereas the dominant contribution of the real part is the off-diagonal. Adding extra poles improves the fit especially for the non-dominant parts of the hybridization function, as one would expect. This is exemplified in Fig. \ref{fig:fit_improv}, where we show the fit result for diagonal and off-diagonal terms for the imaginary part of a (2$\times$2) hybridization function as a function of the number of poles.

Eventually, the error of the fit reaches the $10^{-5}$ threshold, which is the value we chose for the truncation of the $X_\ell$ matrices. Thus, after this limit is reached, the SDR method stops adding the maximal number of effective baths per pole (the $X_\ell$ matrices stop being full rank). One can further improve the quality of the fit by decreasing this threshold, at the price of increasing the number of effective baths. Conversely, we can use the SDR method to decide the number of effective baths needed to reach a particular quality of hybridization fit. For the ($4\times4$) case, it seems that before the fit error reaches the $10^{-5}$ threshold the $X_\ell$ matrices stop being full rank.

\subsubsection{Timings}

\begin{table*}[hbt!]
  \begin{center}
    \begin{tabular}{|r||c|c|c|c|c|c||c|c|c|c|c|c|} 
    \hline
      &\multicolumn{6}{c||}{$\mathbf{(2\times2)}$} &\multicolumn{6}{c|}{$\mathbf{(3\times3)}$ } \\
      \hline
       & \multicolumn{2}{c|}{\textbf{SDR}} & \multicolumn{2}{c|}{\textbf{BFGS}} & \multicolumn{2}{c||}{\textbf{BOBYQA}} & \multicolumn{2}{c|}{\textbf{SDR}} & \multicolumn{2}{c|}{\textbf{BFGS}} & \multicolumn{2}{c|}{\textbf{BOBYQA}}\\
      \hline
      \textbf{$N_p$} & [s] & Error & [s] & Error & [s] & Error & [s] & Error & [s] & Error & [s] & Error\\
      \hline
         2  &  5 & 2.18e-3 &   4 & 1.09e-1 &   28 & 2.18e-3 &   8 & 2.16e-3 &  87 & 9.24e-2 & 19729 & 1.17e-2\\   
                 3  & 17 & 1.59e-3 &   5 & 9.98e-2 &  989 & 1.47e-3 &  16 & 1.57e-3 & 183 & 2.72e-2 & 21429 & 1.15e-2\\   
                 4  &  8 & 2.88e-4 &  33 & 4.90e-3 & 2717 & 1.29e-4 &  14 & 3.97e-4 & 262 & 5.10e-3 & 31889 & 1.15e-2\\   
                 5  &  7 & 2.64e-4 &  48 & 3.30e-3 & 5535 & 1.43e-4 &  24 & 2.93e-4 & 363 & 1.50e-3 & 40528 & 1.15e-2\\   
                 6  & 27 & 1.75e-5 &  56 & 3.43e-4 & 1060 & 1.14e-4 &  29 & 5.99e-5 & 458 & 1.20e-3 & $> 24$h & --\\   
                 7  & 46 & 1.14e-5 &  69 & 2.27e-4 & 1629 & 9.64e-5 &  68 & 4.04e-5 & 522 & 1.10e-3 & 75922 & 1.15e-2\\   
                 8  & 28  & 2.50e-6 &  79 & 9.92e-5 & 1790 & 7.83e-5 & 102 & 5.41e-5 & 617 & 1.10e-3 & $> 24$h & --\\   
                 9  & 96 & 1.56e-6 &  87 & 9.87e-5 & 2633 & 5.35e-5 &  56 & 5.26e-5 & 651 & 1.10e-3 & 45882 & 1.15e-2\\   
                10  & 39  & 1.50e-6 & 113 & 6.09e-5 & 1844 & 7.41e-5 & 129 & 3.08e-5 & 798 & 1.10e-3  & 84600 & 1.15e-2\\   
                \hline
    \end{tabular}
  \end{center}
  \caption{Timings in seconds and final fit errors for the hybridization function fit using the SDR, quasi-Newton (BFGS), and derivative-free (BOBYQA) methods as a function of the number of poles $N_p$. When using BFGS or BOBYQA, the number  of baths $N_b$ corresponds to $N_p$ times the number of sites in the cluster boundary $N^{'}_c$. The same set of initial bath energies were used for all fits. The BFGS method was allowed to perform 2000 iterations.}
  \label{Table:timings}
\end{table*}

To show the superior timing of the SDR fitting procedure, we compare it to the BFGS algorithm~\cite{Broyden1970} as supported by the \texttt{fminunc} routine in MATLAB and the BOBYQA method (a derivative-free optimization method) implemented  in the nlopt library \cite{bobyqa, nlopt} applied directly to Eq.~\eqref{eq:optimization}. The timings in seconds for a number of SDR, BFGS, and BOBYQA fits are reported in Tab.~\ref{Table:timings} for different numbers of effective baths in both the (2$\times$2) and (3$\times$3) clusters. The number of baths used for the BFGS and BOBYQA optimizations is equal to $N_b = N_p\cdot N^{'}_c$ ($N^{'}_c$ being the number of boundary sites in the cluster), which is the maximal number of effective baths used for the SDR with a given fixed number $N_p$ of poles. Additionally, we report the final errors of all three methods. To provide a fair comparison, all fits were started with the same initial guesses for the bath energies and had the same error convergence tolerance of $10^{-5}$. For the initial pole energies, we chose values symmetrically distributed around zero in a spread of $\pm10\ t$, e.g. in the $N_p = 2$ case we chose $\epsilon_p = \pm 2\ t$. In the cases with odd $N_p$ we set one of the initial poles at 0 t. Each of these poles corresponded to $N^{'}_c$ baths of the same energy for the BFGS and BOBYQA methods. The BFGS optimizations were limited to 2000 iterations. Increasing this number would eventually improve the error at the cost of longer runs, but we comment that in practice the possiblity for further improvement is not substantial.

From Table~\ref{Table:timings}, the advantage of the SDR fit becomes clear. It is in general faster and, in the large bath limit, more effective than both BFGS and BOBYQA. For the small $N_p$ values, i.e. $N_p \le 4$, BOBYQA seems to be slower than BFGS, but for the ($2\times2$) cluster it reaches a fit quality at least as good as the SDR fit. The BFGS optimizations in the ($2\times2$) cluster with $N_p = 2,3$ (i.e., 8 and 12 baths, respectively) converge to an error almost two orders of magnitude larger than the errors for SDR and BOBYQA. They are stuck in local minima. The rest of the BFGS calculations were stopped after performing 2000 iterations. It is important to note that we are not claiming that the errors reported in Table~\ref{Table:timings} represent the best possible fit that the BFGS or BOBYQA algorithm can provide for the problem at hand. In fact, a clever choice of the initial guess and exploitation of symmetries presented in the system can greatly improve the optimization. The important conclusion to draw is that when used plainly, without such case-specific improvements, the SDR method clearly outperforms common optimization routines in the hybridization fitting problem. In what follows we show that the SDR fit is empirically robust with respect to these specializations. 

\subsubsection{Robustness to initial guess}
\label{sec:res_fit_rob}

\begin{figure*}
\begin{subfigure}[t]{0.5\textwidth}
  \includegraphics[width=\linewidth]{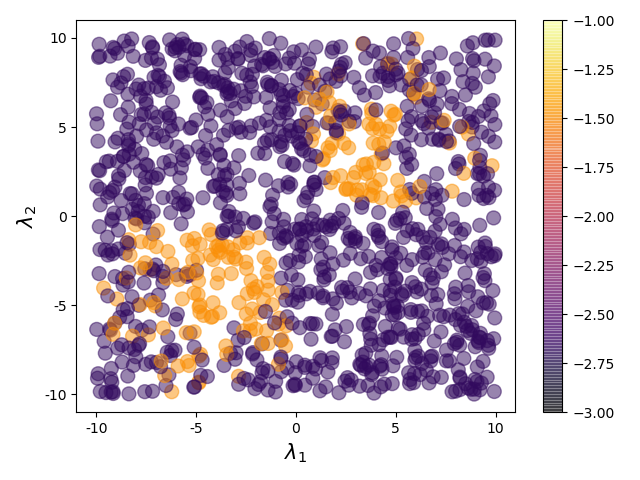}
  \caption{$N_p = 2$, fit error vs. initial pole energies.\\  The best fits have the darkest color.}
  \label{fig:2pole_scatter}
\end{subfigure}%
\begin{subfigure}[t]{0.5\textwidth}
  \includegraphics[width=\linewidth]{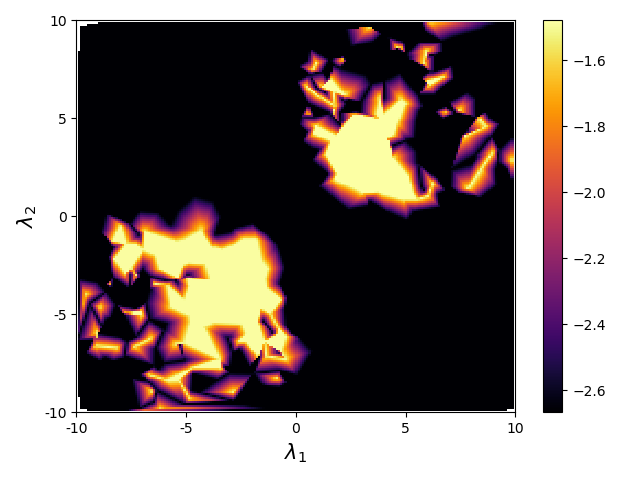}
  \caption{$N_p = 2$, fit error vs. initial pole energies, linear interpolation.  The best fits are colored black.}
  \label{fig:2pole_interp}
\end{subfigure}
\begin{subfigure}[t]{0.5\textwidth}
  \includegraphics[width=\linewidth]{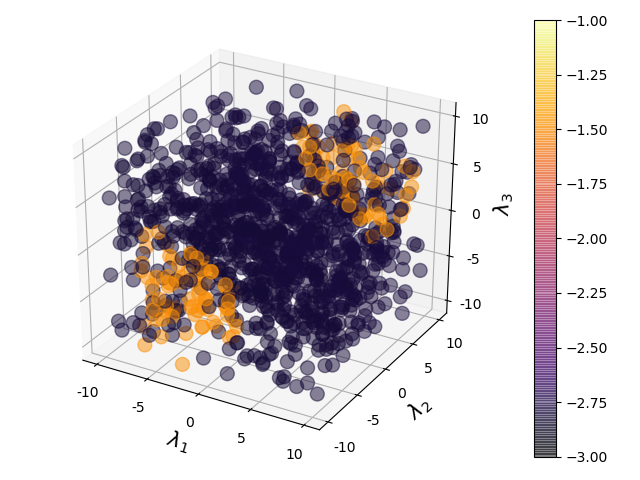}
  \caption{$N_p = 3$, fit error vs. initial pole energies.}
  \label{fig:3pole_scatter}
\end{subfigure}%
\begin{subfigure}[t]{0.5\textwidth}
  \includegraphics[width=\linewidth]{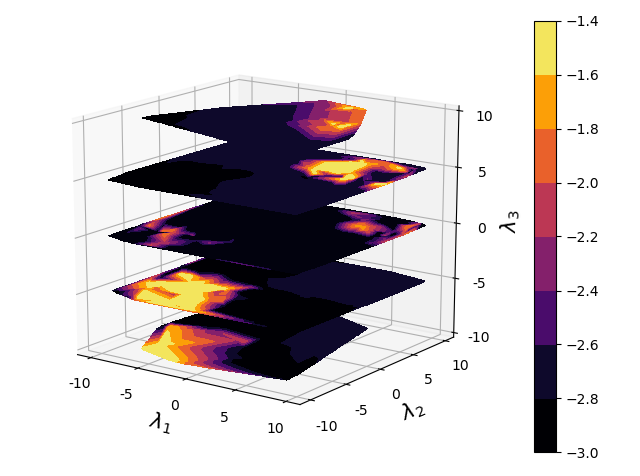}
  \caption{$N_p = 3$, fit error vs. initial pole energies, linear interpolation.}
  \label{fig:3pole_interp}
\end{subfigure}
\caption{Logarithm of the fitting error for a ($3\times3$) hybridization function using $N_p = 2$ (panels~\ref{fig:2pole_scatter} and~\ref{fig:2pole_interp}) and $N_p = 3$ (panels~\ref{fig:3pole_scatter} and~\ref{fig:3pole_interp}) vs. the initial pole energies. Presented are results for over 1000 initial points for each $N_p$, with the initial poles being drawn randomly from a uniform distribution. We report both the actual data as scatter plots and linearly interpolated heat maps. See text for details.}
\label{fig:robust_points}
\end{figure*}

\begin{figure*}
\begin{subfigure}[t]{0.5\textwidth}
  \includegraphics[width=\linewidth]{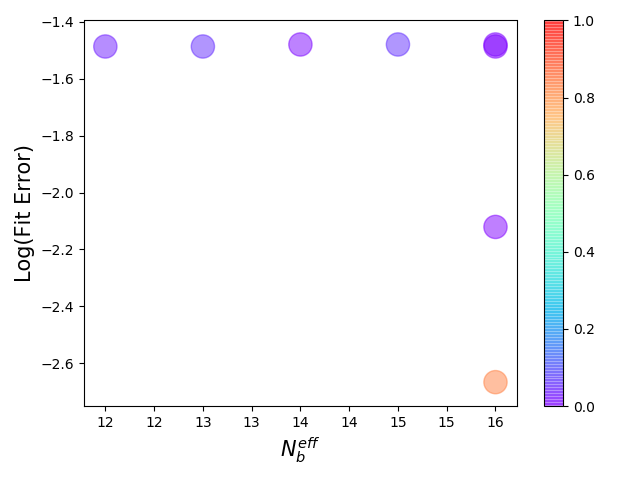}
  \caption{$N_p = 2$.}
  \label{fig:2poles_effBath}
\end{subfigure}%
\begin{subfigure}[t]{0.5\textwidth}
  \includegraphics[width=\linewidth]{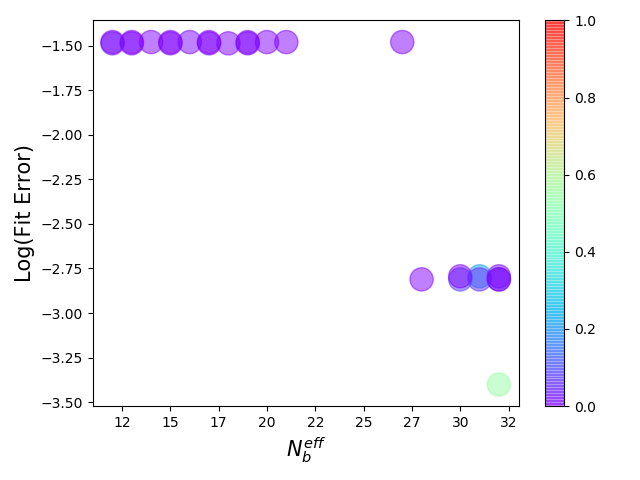}
  \caption{$N_p = 4$.}
  \label{fig:4poles_effBath}
\end{subfigure}
\caption{Logarithm of the fitting error for a ($3\times3$) hybridization function using $N_p = 2$ (left) and $N_p = 4$ (right) vs. the effective bath size $N^{\text{eff}}_b$. Presented are results for over 1000 initial points for each $N_p$, with the initial poles being drawn randomly from a uniform distribution. The color-code corresponds to the relative occurrence of each (error, $N^{\text{eff}}_b$) pair. See text for details.}
\label{fig:robust_effBath}
\end{figure*}

The goal of DMFT is to describe a many-body system, with particular interest in determining phase diagrams. When there are several competing orders this is a highly non-trivial task. In particular, for an iterative self-consistent method like DMFT, it is important to avoid falling into unphysical fixed points of the self-consistent loop. The fit plays a very important role here, and it is desirable for it to find the right set of bath parameters as independently as possible from the initial guess parameters and the frequency grid.

Here, we want to show empirically that the SDR fit is robust with respect to the initial guess. The effect of other parameters, like the frequency grid or the imposition of symmetries, is discussed in section~\ref{sec:res_DMFT}. While different initial guesses can result in different pole energies using the SDR fit, for all the impurity models studied in this work, empirically we find it  easy to distinguish between effective and non-effective fits. Over the set of test systems studied in this work, all $X_\ell$ matrices are of full rank in the case of the best fits. On the other hand, when the SDR method provides $X_{\ell}$ matrices of smaller rank, we are able to change the initial guess and find a better fit with full rank matrices, which reduces the error by one order of magnitude.
Hence the robustness of the SDR fit is evidenced by the empirical realization that all effective fits for a given hybridization function had numerically equal pole energies and bath couplings. In other words, all initial parameters that produced an effective fit, i.e. which maximized the ranks of the $X_\ell$ matrices, resulted in the same pole energies and couplings. It is this remarkable coincidence which makes us confident, at least empirically, of the robustness of the method. To the best of our knowledge, this property sets the SDR fit apart from all other non-linear optimization methods used for fitting of impurity models.

\begin{table}[hbt!]
  \begin{center}
    \begin{tabular}{|l|c|c|c|} 
    \hline
    & $\mathbf{N_p=2}$ & $\mathbf{N_p=3}$ & $\mathbf{N_p=4}$  \\
    \hline
    Nr. of samples & 1184 & 1195 & 1469 \\
    Rate of Success & $83\%$ & $88\%$ ($43\% + 45\%$) & $55\%$\\
    max[$\sigma(\epsilon_\ell)$] & 6e-6 & 5e-4 & 2e-4\\
    \hline
    \end{tabular}
  \end{center}
  \caption{Statistical analysis of the robustness to the initial pole energies for the SDR fit on a $(3\times3)$ cluster hybridization function. Reported are the number of different initial conditions sampled, represented graphically in Fig.~\ref{fig:robust_points}. The different initial pole energies were drawn randomly from a uniform distribution. The rate of success is defined as the number of optimal fits divided by the number of samples. The distribution of the errors for the non-optimal fits are presented in Fig.~\ref{fig:robust_effBath}. In the case of $N_p = 3$, there were two optimal sets of bath parameters, occurring approximately with a 1:1 probability, see text for details. The last row reports the maximal standard deviation of each of the pole energies in the optimal fit parameters. This represents the spread of the pole energies in the optimal fit parameters.}
  \label{Table:robustness}
\end{table}

We exploited this property when choosing the initial set of pole energies for the input of the SDR fit. Originally, we would choose random energies from a normal distribution with average at negative chemical potential $-\mu$. While this initial guess almost always results in an effective fit, for some examples, particularly with dense frequency grids or with $N_p > 8$ in Fig.~\ref{fig:2x2_energ_conv}, this kind of initial guess could lead to a poor fitting result. In these cases, we chose initial pole energies symmetrically distributed around zero in a spread from $\pm8\ t$. This always achieves an effective fit. In each case where this was necessary, we would test at least three different such initial guesses to assure the robustness of the fit parameters. We achieve an effective fit for most of our practical initializations and ineffective fits can be fixed easily with small modifications to the initialization.

We illustrate this property in Fig.~\ref{fig:robust_points} and~\ref{fig:robust_effBath}, and in Table~\ref{Table:robustness}. In Fig.~\ref{fig:robust_points} we present the logarithmic error of the fit for different values of initial poles in $N_p = 2$ (top) and $N_p = 3$ (bottom), for over 1000 different initial poles uniformly distributed in the $[-10;10]^{\otimes N_p}$. The data suggests that most of the initial guesses result in an optimal fit (the exact proportion is given in Table~\ref{Table:robustness}). The initial guesses that fail are those that have no spread in the energy of the poles, i.e. they are concentrated in the corners of the square (cubic) interval. In Fig.~\ref{fig:robust_effBath} we report the logarithm of the final error vs the obtained  effective bath size $N^{eff}_b$ for $N_p = 2, 4$, also for over 1000 different initial poles each. The color-code corresponds to the relative occurrence of each (error, $N^{\text{eff}}_b$) pair. First, the optimal fit, significantly better than the next best one, is obtained always with the maximal $N^{eff}_b$ and the highest occurrence. While there exist fits of maximal effective bath size and non optimal error, these occur less than 5$\%$ of the times. For the $N_p = 2$ case the optimal fit is found with over 80$\%$ probability, in the $N_p = 4$ case with $55\%$. Thus, further increasing the number of poles probably makes randomly finding an effective starting guess more difficult, but the results in Fig.~\ref{fig:robust_points} suggest that an initial guess with some degree of spreading between the pole energies will likely give a good result. Even if that were not the case, the efficient timings already presented in Table~\ref{Table:timings} suggest that it is always possible to perform some initial statistics in the first DMFT iteration, trying several initial pole energies in a reasonable time. This should be enough to effectively find a good fit. Since in the following iterations one can use the previous fit result as initial guess, the fit should be more likely to converge to the optimal solution in those cases.
 
\begin{table*}[hbt!]
  \begin{center}
    \begin{tabular}{|r||c|c||c|c|} 
    \hline
      & \multicolumn{2}{c||}{$\mathbf{(2\times2)}$} & \multicolumn{2}{c|}{$\mathbf{(3\times3)}$} \\
      \hline
      \textbf{$N_p$} & \textbf{Error Reduction} & \textbf{Time [s]} & \textbf{Error Reduction} & \textbf{Time [s]} \\
      \hline
         2  &   1.0   &    0.4  & 1.2 & 14 \\   
                 3  &   1.0   &    2    & 1.2 & 78 \\   
                 4  &   2.9   &    3    & 3.6   & 96 \\   
                 5  &   4.1   &    17   & 3.6   & 104 \\   
                 6  &   1.1   &    1    & 1.5 & 148 \\   
                 7  &   4.2   &    25   & 1.2 & 151 \\   
                 8  &   1.4 &    2    & 2.0   & 195\\   
                 9  &   1.1   &    1    & 2.1   & 219 \\   
                10  &   1.0   &    1    & 1.5 & 223 \\   
                \hline
    \end{tabular}
  \end{center}
  \caption{Error reduction (multiplicative improvement) and timing for optimizations using the SDR fit and a consecutive BFGS step. The timings refer only to final BFGS optimization.}
  \label{Table:SDRBFGS}
\end{table*} 

Table~\ref{Table:robustness} summarizes the statistics of the robustness tests for $N_p = 2, 3$ and 4. The rate of success, defined as the ratio between the number of times the optimal fitting parameters were found and the number of attempts, decreases drastically from $N_p = 3$ to $N_p = 4$, but is still reasonably high for a random set of initial guesses. The most remarkable fact is still the uniqueness of the optimal solution, as discussed in the previous paragraphs. All optimal fits converge to the same pole energies, with minimal standard deviations as reported in Table~\ref{Table:robustness}. The case with odd $N_p$ is an exception, in which there are actually two optimal sets of pole energies. These are approximately related to each other by a sign change. Such a sign relation is the symmetry suggested by the converged baths in Fig.~\ref{fig:energ_conv} and~\ref{fig:energ_conv_DMFT}, so the existence of this pair of optimal fit parameters for $N_p = 3$ may well be an even-odd artifact rather than an actual feature of the SDR fit.

\subsubsection{Further optimization on top of SDR}

We want to point out that the error achieved by the SDR fit is not necesarily the global minimum, in the sense that it can be improved further among hybridization functions of the form of Eq.~\eqref{eq:ansatz} with a fixed number $N_b = N_p \cdot N_c$ of baths. By simply breaking the pole degeneracy, one can further reduce the fitting error. After having performed an SDR fit, the bath parameters obtained can then be used as starting guesses for a further optimization procedure, this time using all bath energies and couplings as free parameters at the same time. We used the BFGS algorithm and were able to improve the error by a factor between 1 and 4 depending on the number of poles $N_p$ and the size of the cluster. Performing this gradient-based optimization on top of the SDR increases the total fitting time by an amount which depends on the number of baths and the cluster size. We present the factor of error improvement and timing for different fitting problems in Table \ref{Table:SDRBFGS}.

\begin{figure}
\includegraphics[width=0.5\textwidth]{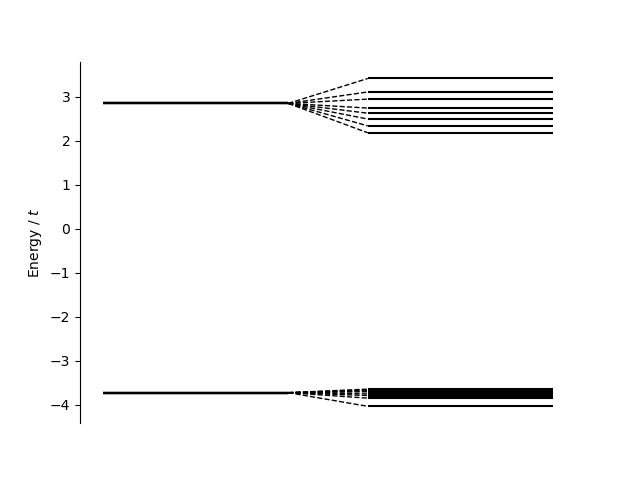}
\caption{Pole energies computed by the SDR fit for a (3$\times$3) cluster hybridization with $N_p = 2$ before (left) and after (right) an additional BFGS optimization. Since the BFGS step does not maintain the degenerate pole structure, instead using all bath energies and couplings independently and simultaneously, the pole energies spread. See text for details.}
\label{fig:BFGSspread}
\end{figure}

Considering the timings and factors of improvement in Table \ref{Table:SDRBFGS}, it is not always advantageous to perform the additional optimization, and one should consider whether the small error reduction is worth the computational cost. The degree to which the pole energies spread out after a gradient-based step is shown in Fig. \ref{fig:BFGSspread} on the example of an SDR fit with $N_p=2$ for a (3$\times$3) hybridization function.

\subsubsection{Pole energy convergence}

\begin{figure*}
\begin{subfigure}[t]{0.5\textwidth}
  \includegraphics[width=\linewidth]{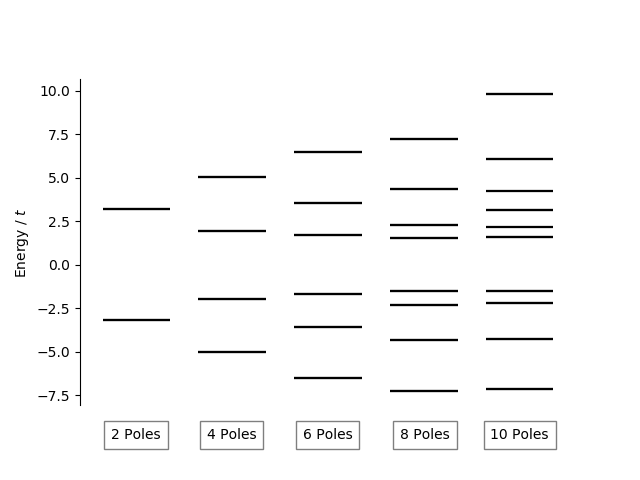}
  \caption{(2$\times$2) cluster.}
  \label{fig:2x2_energ_conv}
\end{subfigure}%
\begin{subfigure}[t]{0.5\textwidth}
  \includegraphics[width=\linewidth]{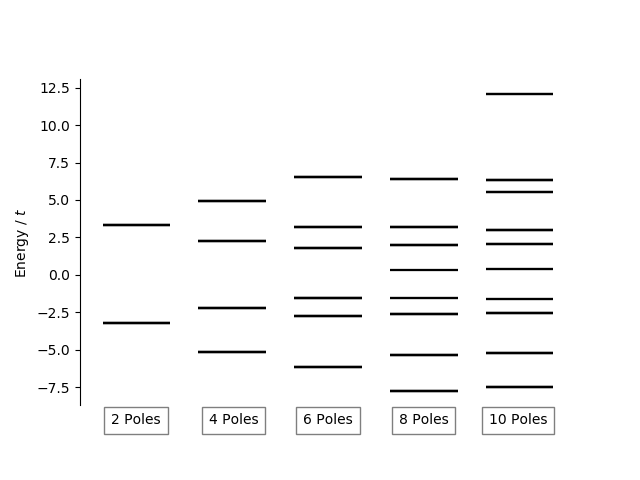}
  \caption{(3$\times$3) cluster.}
  \label{fig:3x3_energ_conv}
\end{subfigure}
\caption{Pole energies resulting from the SDR fits for the (2$\times$2) (left) and (3$\times$3) (right) clusters for $N_p = 2,4,6,8,10$. We see that the pole energies are fairly symmetric around zero energy and that the low energy poles converge very quickly. See text for details.}
\label{fig:energ_conv}
\end{figure*}

In the previous subsections we have characterized in detail the important numerical properties of the SDR fit, independently of its influence on the full DMFT loop. Before describing how the SDR fit can improve the DMFT iteration, there is a final convergence property, of particular importance to justify the bath truncation in Hamiltonian-based DMFT, that we would like to address: the convergence of the pole energies with respect to the number of poles. To this end, in Fig. \ref{fig:energ_conv} we report the pole energies for the (2$\times$2) and (3$\times$3) hybridizations as a function of the number $N_p$ of poles used in the SDR fit. (For simplicity, we only report even numbers of poles.) The pole energies seem to be symmetric around the zero energy for both the (2$\times$2) and (3$\times$3) clusters for $N_p = 2, 4, 6$. It should be noted that this is not a constraint on the SDR fit and that this symmetry is found numerically by the method without imposing explicitly any symmetry constraint. The``emergence" of other such symmetries during the DMFT loop with the SDR fit will be discussed in detailed in the section~\ref{sec:res_DMFT}. 

Fig. \ref{fig:energ_conv} suggests that the low energy poles converge quickly. By this we mean that adding extra poles does not significantly alter the energies of the previous ones. Effectively this means that that even fits with relatively few poles can describe the low energy properties of the embedded system and that additional poles eventually only play a role for the simulation of the high energy features. This feature is of great importance for Hamiltonian-based implementations of DMFT, where a truncation of the bath is necessary. The results in Fig. \ref{fig:energ_conv} effectively corroborate the physical intuition that such a bath truncation is a viable approximation when one is only interested in describing the low energy features of the model. It should be noted that these are results for fits on a single hybridization function. Physically, one would need to converge a full DMFT calculation for each number of poles (each bath size) and compare those converged pole energies to make a definitive statement about the energy convergence in the embedded model. We present such a study in the following section, where the performance of the SDR fit inside full cluster DMFT calculations is presented in detail.

\subsection{Performance in DMFT calculation}
\label{sec:res_DMFT}


Having characterized the performance of the SDR fit for impurity problems in detail, we now turn to showing how its properties affect full cluster DMFT calculations. There are four points that we would like to stress here, beyond those already presented in the previous section: 

\begin{enumerate}
    \item The pole energy convergence, which we have demonstrated for a single hybridization fit, translates into convergence of the bath energies of fully converged DMFT calculations, as the number of poles is increased.
    \item The SDR fitting procedure finds, up to a negligible numerical error, symmetries present in the system without a need for imposing them.
    \item Given a large enough number of poles, the SDR fit result is largely unaffected by the imaginary frequency range used to sample the hybridization function.
    \item Given its timing and flexibility, the SDR fit works for both highly symmetric clusters where a block-diagonalization of the hybridization function is possible~\cite{Koch2008} and those clusters where this is not feasible.
\end{enumerate}

As in the previous section, all calculations presented here are based on a two-dimensional square-lattice one-band Hubbard model at half-filling. 

\begin{figure*}
\begin{subfigure}[t]{0.5\textwidth}
  \includegraphics[width=\linewidth]{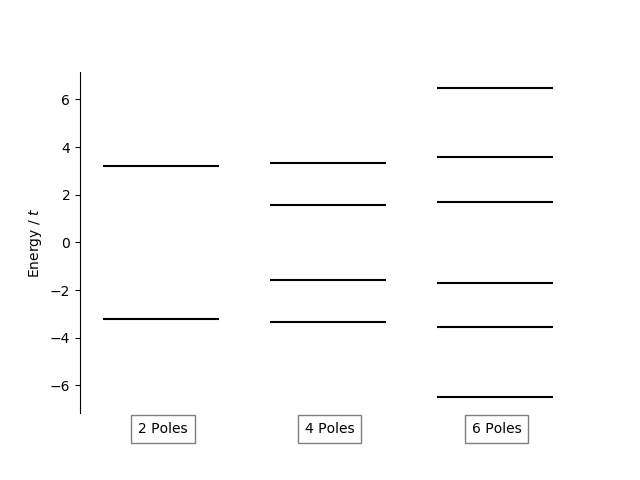}
  \caption{(2$\times$2) cluster.}
  \label{fig:2x2_energ_conv_DMFT}
\end{subfigure}%
\begin{subfigure}[t]{0.5\textwidth}
  \includegraphics[width=\linewidth]{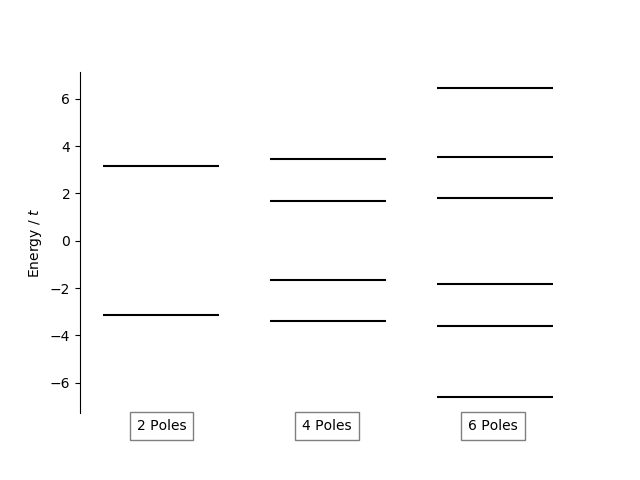}
  \caption{(2$\times$3) cluster.}
  \label{fig:2x3_energ_conv_DMFT}
\end{subfigure}
\begin{subfigure}[t]{0.5\textwidth}
  \includegraphics[width=\linewidth]{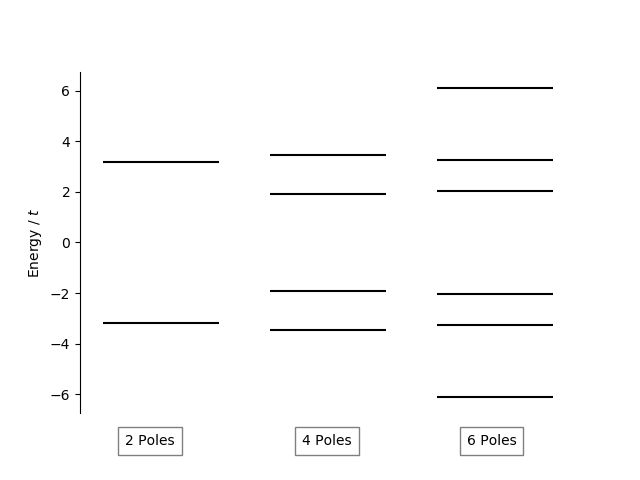}
  \caption{(2$\times$4) cluster.}
  \label{fig:2x4_energ_conv_DMFT}
\end{subfigure}%
\begin{subfigure}[t]{0.5\textwidth}
  \includegraphics[width=\linewidth]{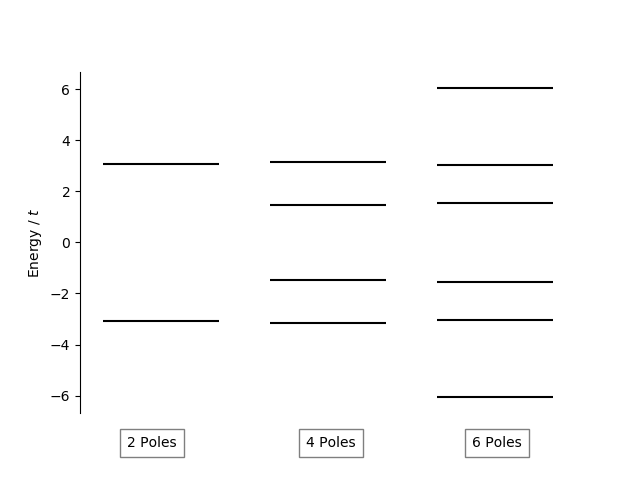}
  \caption{(1$\times$6) cluster.}
  \label{fig:1x6_energ_conv_DMFT}
\end{subfigure}
\caption{Pole energies computed by the SDR fit for clusters of different size with $N_p = 2,4,6$. These are converged in cluster DMFT calculations at $U\ / \ t = 8$.}
\label{fig:energ_conv_DMFT}
\end{figure*}

We now expound upon the advantage of the SDR fit in DMFT for clusters of low symmetry. The calculations of the previous section treat highly symmetric clusters, in particular square clusters embedded in the full square lattice. For these symmetric clusters, one can usually block-diagonalize the hybridization function, making the fitting problem much easier~\cite{Koch2008,Liebsch2011}. For less symmetric clusters, however, this is not necessarily possible, and a flexible and efficient fitting method like SDR is important. To show that the SDR fit can handle lower-symmetry clusters with the same ease as fully symmetric ones, we present in this section  results from DMFT calculations in (2$\times$3), (2$\times$4) and (1$\times$6) clusters of the two-dimensional square-lattice Hubbard model at half-filling and $U\,/ \, t = 8$. Being able to treat a pseudo-one-dimensional cluster, or stripe, like the (1$\times$6) cluster opens up interesting avenues of research, in light of the ongoing quest for stripe order in the under doped Hubbard model~\cite{Hager2005,Zheng2017,Huang2017,Huang2018}. All converged bath parameters used to produce the data in this section are collected in tables in the supporting information.

\subsubsection{Convergence of the bath energy}

In the previous section we showed that, for a single hybridization fit, the low energy baths converge rapidly with increasing number of poles (see Fig.~\ref{fig:energ_conv}) which is an important property for the bath truncation in Hamiltonian-based DMFT approaches. The results presented in Fig.~\ref{fig:energ_conv} originated from one single hybridization fit, and were computed using numbers of poles well beyond the capabilities of state-of-the-art zero temperature impurity solvers. It is fundamental  to test the same convergence behavior when considering fully converged cluster DMFT calculations for the bath sizes that can be handled. Fig.~\ref{fig:energ_conv_DMFT} shows the converged bath energies for different cluster DMFT calculations on (2$\times$2), (2$\times$3), (2$\times$4) and (1$\times$6) clusters with $N_p = 2,4,6$. The initial guess for the poles in the fits DMFT loop was chosen in the same way described in section~\ref{sec:res_fit_rob}, i.e. by choosing random pole energies centered around the negative chemical potential $-\mu$. For the subsequent DMFT iterations, the previous pole energies were used as initial guess. Numerical convergence of the pole energies was obtained after 10 to 20 DMFT iterations.

\begin{figure*}
\includegraphics[]{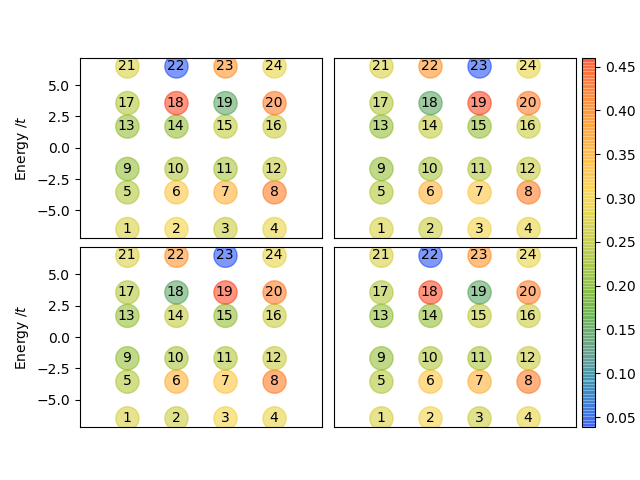}
\caption{Absolute value of the bath couplings $|V_{\alpha, \ell}|$ for a (2$\times$2) cluster DMFT calculation with 24 baths ($N_p = 6$). Each subplot corresponds to one site in the (2$\times$2) cluster, the circles representing the different bath sites. The energy of each bath site is represented on the y-axis, while the color encodes the absolute value of the coupling $| V_{\alpha,\ell}|$ between the baths and the cluster sites.
}
\label{fig:2x2_24_couplings}
\end{figure*}

\begin{figure*}
\includegraphics[]{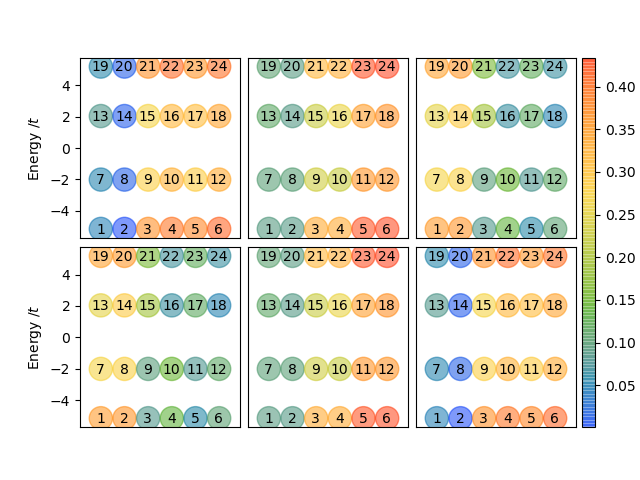}
\caption{Absolute value of the bath couplings $|V_{\alpha, \ell}|$ for a (2$\times$3) cluster DMFT calculation with 24 baths ($N_p = 4$). Each subplot corresponds to one site in the (2$\times$3) cluster, the circles representing the different bath sites. The energy of each bath site is represented on the y-axis, while the color encodes the absolute value of the coupling $|V_{\alpha, \ell}|$ between the baths and the cluster sites.}
\label{fig:2x3_24_couplings}
\end{figure*}

The most apparent property is the near-perfect symmetry of the bath energies around 0 in both cases. This is not explicitly imposed in the SDR procedure and may be a consequence of the particle-hole symmetry present in the one-band Hubbard model at half filling. As a consequence, calculations with an odd number of poles recovered one bath energy close to zero. The authors have not observed this behavior with other fitting methods, like the BOBYQA optimization~\cite{nlopt,bobyqa}. Meanwhile, the conclusion from the previous section still holds: the low-lying pole energies converge rapidly as the number $N_p$ of poles is increased. We have omitted the case of odd $N_p$ since there are intrinsic even-odd disparities that do not reflect the actual convergence properties of the pole energies.
While the impurity solver imposes a strict limit on the total number of bath sites (in particular, preventing exploration beyond $N_p=6$), it seems fair to say that the two lowest-lying pairs of baths have converged their energies at $N_p = 6$, and probably adding further poles will only yield higher-energy baths. The positive interpretation for the validity of bath truncation schemes remains.

\subsubsection{Emergence of symmetry}

Imposing the symmetries present in the cluster in the fitting process is a necessary step with most currently employed optimization schemes. In such schemes, this step both simplifies the fit and assures that the right physical fixed point in the DMFT self-consistent cycle is obtained~\cite{Koch2008, Liebsch2011}. While it is possible to adopt these strategies when using the SDR approach, we have observed that this is not necessary since the bath couplings $V_{\alpha,\ell}$ computed using the SDR fit already show symmetry, at least qualitatively. By this we mean that all symmetrically equivalent cluster sites experience equivalent baths. To illustrate this point, in Fig.~\ref{fig:2x2_24_couplings} we report the absolute value of the bath couplings for all cluster and bath sites of a (2$\times$2) cluster DMFT calculation with 24 baths ($N_p = 6$) . In this figure, the (2$\times$2) cluster is represented by the (2$\times$2) grid of subplots, each subplot corresponding to one of the cluster sites. The circles then correspond to the 24 baths, their energies shown in their y-axis position and their absolute value of the coupling $|V_{\alpha, \ell}|$ encoded in the color map.

From Fig.~\ref{fig:2x2_24_couplings} it becomes apparent that all four cluster sites experience an equivalent bath. For each given bath energy, the four cluster sites couple to baths that present the same four absolute coupling amplitudes, albeit in a different order depending on the cluster site. This permutation is physically inconsequential however, since the bath energies are degenerate. The full coupling values $V_{\alpha,\ell}$ can differ by a sign, but it is only their amplitude square, understood as a transition probability in a Fermi-golden-rule argument, that bears physical meaning. We want to stress that we do not explicitly impose any symmetry constraint on the SDR fit and that it arrives at this symmetric bath configuration automatically. The only user-provided input is the initial guess for the bath parameters for the first DMFT iteration, which is chosen by drawing random numbers around the chemical potential $\mu$ for the pole energies and between 0 and 1 for the bath couplings. For this initial bath guess, we do impose symmetry by assigning the same bath couplings to symmetrically equivalent cluster sites. Nonetheless, this aspect of the initial guess only enters via the computation of the Hamiltonian; the SDR uses a random initial guess for the bath couplings. Furthermore, the SDR procedure relaxes the symmetries of the initial bath guess: the bath couplings to symmetrically equivalent sites are not identical, but merely equivalent up to the permutations as previously discussed and demonstrated in Fig.~\ref{fig:2x2_24_couplings}. We have not observed such symmetry emergence from other fitting methods like the BOBYQA algorithm.

\begin{figure*}
\includegraphics[]{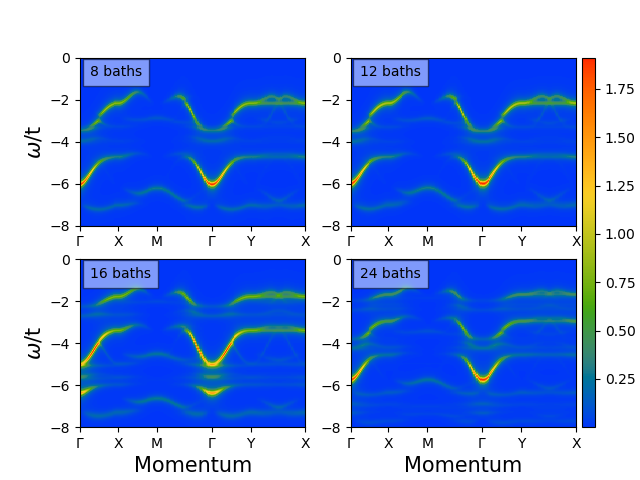}
\caption{Spectral weights for (2$\times$2) cluster DMFT calculations on the square-lattice one-band Hubbard model at half-filling with $U \, / \, t = 8$ and different numbers of baths. SDR fit performed using a linear imaginary frequency grid with frequency cutoff at $i\omega_{max} = 40\ t$.}
\label{fig:akw_2x2_oldFreq}
\end{figure*}

Such behavior is not limited to the fully symmetric (2$\times$2) cluster, but in fact also holds for rectangular clusters that do not enjoy all the symmetry of the full square lattice. To demonstrate this, we report in Fig.~\ref{fig:2x3_24_couplings} the absolute values for the couplings of a (2$\times$3) cluster with 24 baths ($N_p = 4$). Here, an interesting thing occurs. From the full symmetry group of a (2$\times$3) slab, one would expect to only have two symmetrically inequivalent sites: the four corners and the two center sites. This is due to the $C_2$ axis and the two reflection planes perpendicular to the cluster plane. This observation notwithstanding, as can be seen in Fig.~\ref{fig:2x3_24_couplings}, the SDR fit seems to break the reflection symmetries and identify three different types of sites, leaving only the $C_2$ perpendicular rotation symmetry. The small number of poles may cause the SDR routine to break the symmetry in order to improve the fit. It is plausible that increasing the number of poles would remedy this loss of symmetry, but the large number of bath degrees of freedom that this analysis would need exceeds the limitations of most impurity solvers.

\subsubsection{Effect of the imaginary frequency grid}

\begin{figure*}
\includegraphics[height=0.35\textheight]{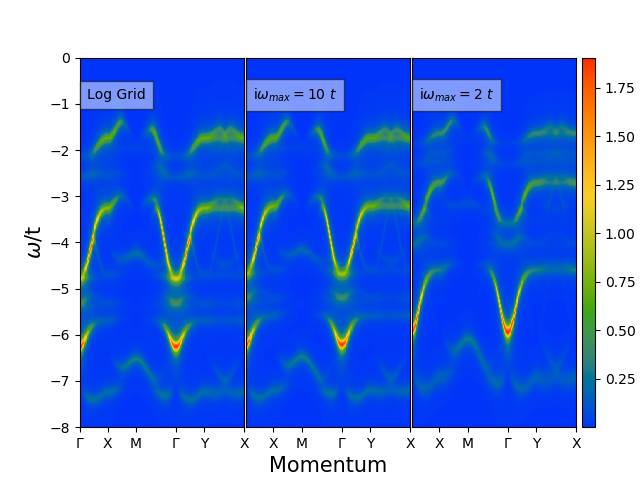}
\caption{Spectral weights for (2$\times$2) cluster DMFT calculations with 16 baths ($N_p = 4$) on the square-lattice one-band Hubbard model at half-filling with $U\, / \, t = 8$. SDR fit performed using different imaginary frequency grids (see text for details).}
\label{fig:akw_2x2_diffGrids}
\end{figure*}

The fact that the DMFT self-consistency loop is performed along the imaginary frequency axis has both advantages and disadvantages. On the one hand, it simplifies the calculations and fits, since the Green's function is smooth along the imaginary axis and decays as~$\propto \left| \frac{1}{\omega} \right|$. On the other hand, it increases the difficulty of extracting physical information contained in Green's functions near the real axis. Indeed, the real-frequency physical details of the system, e.g., the existence or absence of a gap, are encoded close to the real axis with small imaginary frequency values, and the long frequency tail mainly includes information about sum rules. To understand this last point, it suffices to consider that the Green's function along the imaginary frequency axis depends on its value on the real axis according to

\begin{equation} \label{eq:ImagFreq}
        G(i\omega_n) = -\frac{1}{\pi}\int_{-\infty}^{\infty}\mathrm{d}\omega\ \frac{\mathrm{Im}(G(\omega))}{i\omega_n-\omega}.
\end{equation}
Sum rules are written as integrals over the full real frequency axis, e.g., the relation
\begin{equation} \label{eq:SimpleSumRule}
        -\frac{1}{\pi}\int_{-\infty}^{\infty}\mathrm{d}\omega\ \mathrm{Im}(G(\omega)) = 1.
\end{equation}
Such non-local information in the frequency space is stored across the entire imaginary frequency axis, and to recover it one needs to consider the large-frequency regime as well as the low-frequency one. This is manageable at finite temperature, where the Green's function is only defined on a finite set of imaginary frequencies (the Matsubara frequencies). However, at zero temperature, one needs in principle the full positive imaginary frequency axis to get the full physics, while on the real axis it usually suffices to consider features close to a non-interacting Fermi energy.

Maybe the most immediate and palpable consequence for zero temperature Hamiltonian-based DMFT calculations is the fact that the frequency grid used for the fit can have a large impact on the reliability of the results. How closely this grid approaches the real axis and how far into the high frequency limit it extends can change the qualitative behavior of the real-frequency results. Of course, one cannot make the grid arbitrarily vast and dense, since doing so increases the difficulty of the optimization step. As a consequence, it requires a significant amount of physical intuition about the problem to decide the limits and density of this frequency grid. In some circumstances, it may even be preferable to choose a dense grid, very close to the real axis but ignoring the high frequency limit, to achieve satisfactory results. 

In a previous work~\cite{Mejuto2019}, such behavior was observed using the BOBYQA algorithm for the fit in a series of (2$\times$2) cluster DMFT calculations on the two-dimensional square-lattice one-band Hubbard model at half-filling and $U \, / \, t = 8$, when increasing the number of bath sites. In a series of cluster DMFT calculations with 8, 12, 16 and 24 baths, we observed a drastic change in the high frequency ``hole-bands" of the spectral weight when increasing the number of baths from 12 to 16, and in turn to 24. We argued that the drastic change was due to over-fitting, in the sense that the BOBYQA algorithm was reducing the cost function in Eq.~\eqref{eq:CostFunc} in the high frequency tail at the expense of a poorer fit close to the real axis. However, as discussed above, this is where the details of the physics of the system are encoded. After limiting the fitting frequency grid to the immediate vicinity of the real axis, the drastic change in the large bath limit disappears, and thus we concluded that this change is not physical, but rather an artifact of the fit. Here, we perform a similar study using the SDR fit.

First, for comparison with the results obtained in~\cite{Mejuto2019}, we report in Fig.~\ref{fig:akw_2x2_oldFreq} the spectral weights $A(\mathbf{k},\omega)$, defined as
\begin{equation} \label{eq:Akw}
        A(\mathbf{k},\omega) = -\frac{1}{\pi}\mathrm{Im}(G^p_{latt}(\mathbf{k},\omega)) ,
\end{equation}
for the same (2$\times$2) cluster DMFT calculations described in the previous paragraph (i.e., using $N_p = 2,3,4,6$), using the customary linear frequency grid of 50 points along the imaginary frequency axis from $i\omega = 0\ t$ to $i\omega = 40\, t$ for the DMFT self-consistency. The superscript $p$ in the lattice Green's function $G^p_{latt}(\mathbf{k}, \omega)$ accounts for the fact that one has to recover the translational symmetry broken in the cluster DMFT treatment. In this work we employ the following  periodization scheme~\cite{Go2011,Go2017}
\begin{equation}
G^p_{latt}(\mathbf{k},\omega) = \frac{1}{N_c}\sum_{\alpha,\beta=1}^{N_c}e^{i\mathbf{k}(\mathbf{r}_\alpha-\mathbf{r}_\beta)}G_{latt}(\mathbf{k}, \omega)_{\alpha,\beta},
\label{eq:GFlatt_p}
\end{equation}
where $G_{latt}(\mathbf{k},\omega)$ is the unperiodized lattice Green's function in Eq.~\eqref{eq:GFlatt} evaluated on the real frequency axis, and $\mathbf{r}_\alpha$ are the relative position vectors of the lattice sites inside the cluster. Now Fig.~\ref{fig:akw_2x2_oldFreq} shows the same drastic change in the high-frequency bands when changing from 12 to 16 baths as the one observed in~\cite{Mejuto2019}. However, this change seems to be rectified in the 24 bath calculation. Before getting into this significant distinction between the BOBYQA and SDR results, we want to show explicitly that changing the frequency range indeed gets rid of the strange behavior for the 16 baths ($N_p =4$) calculation.

\begin{figure}
\includegraphics[width=0.5\textwidth]{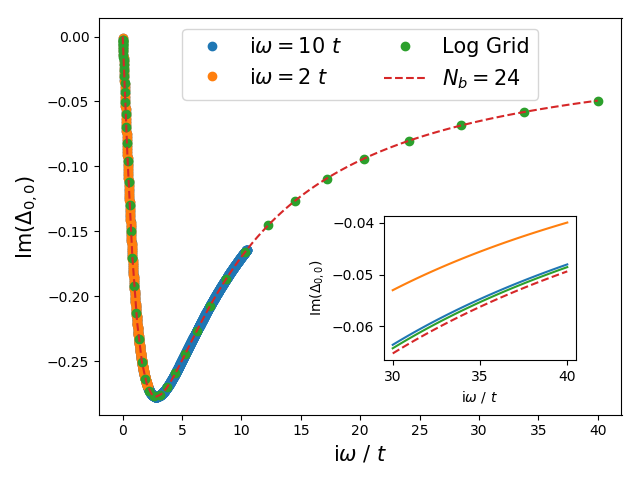}
\caption{Imaginary part of diagonal component of the hybridization function $\Delta$ along the imaginary frequency axis for DMFT calculations on a (2$\times$2) cluster with 16 baths ($N_p = 4$) and three different frequency grids (see text for details) and with 24 baths ($N_p = 6$). The fit with 24 baths was performed on the linear grid with 50 points from $i\omega\ /\ t = 0$ to $i\omega\ /\ t = 40$. The inset shows the same $\Delta$, computed in a dense grid in the high frequency range using the bath parameters of the four different fits.}
\label{fig:diag_hyb_diffFreqs}
\end{figure}

In Fig.~\ref{fig:akw_2x2_diffGrids} we report the spectral weights for the (2$\times$2) cluster DMFT with 16 baths that result from using different frequency grids in the fit. In particular we show a logarithmic grid of 50 points from $i\omega = 0.01\, t$ to $i\omega = 40\, t$ (left panel) and two linear grids of Matsubara frequencies $\omega_n = \frac{(2n + 1)\pi}{\beta}$ with $\beta = 600$, with maximal frequencies of $10\, t$ (center panel) and $2\, t$ (right panel). These three grids were chosen to increase the relative weight of points close to the real frequency axis in the fit, by choosing either the logarithmic spacing or a dense grid of points with a hard cutoff at some small frequency. For reference, the imaginary part of the diagonal component of the hybridization function along the imaginary frequency axis is shown for all these calculations in Fig.~\ref{fig:diag_hyb_diffFreqs}.

From Fig.~\ref{fig:diag_hyb_diffFreqs} it appears that there is no difference between the hybridization functions between the three grids (in fact, the average absolute difference between the two Matsubara grids for $i\omega < 2\, t$ is approximately $2\times 10^{-4}$), and yet the real-axis properties show an appreciable difference (see Fig.~\ref{fig:akw_2x2_diffGrids}). The reason for this is as follows. Computing the hybridization function with bath parameters obtained from the fit which stops at $i\omega = 2\, t$ for imaginary frequencies \emph{beyond} $i\omega = 2\, t$ unveils significant differences between the hybridization functions (see inset in Fig.~\ref{fig:diag_hyb_diffFreqs}). This implies that the high frequency behavior of the hybridization function computed by fitting the smallest of our frequency ranges is not appropriately converged. Indeed, if one adds but five points to the Matsubara frequency range ending at $i\omega = 2\, t$, these points equally distributed from $i\omega = 2\, t$ to $i\omega = 40\, t$, and performs a new DMFT calculation starting from the ``converged" parameters of the small frequency range calculation, the bath parameters rapidly change to improve the high frequency hybridization function. These become equivalent to those from the larger frequency range calculation, or the logarithmic scale one, and the spectral weights recover the structure in the left panel of Fig.~\ref{fig:akw_2x2_diffGrids}. This confirms the fact that to obtain the right physical behavior with a moderate number of baths, there is the need to choose the frequency grid appropriately. This requires some \emph{a priori} knowledge of the solution, and it would be desirable that such fine-tuning would not be necessary. In particular, the hope is that in the large bath limit this bias of the frequency grid becomes unnecessary. 

\begin{figure*}
\includegraphics[]{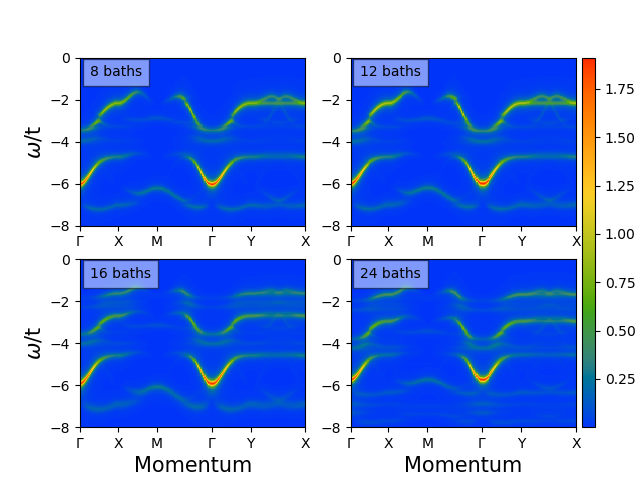}
\caption{Spectral weights for (2$\times$2) cluster DMFT calculations on the square-lattice one-band Hubbard model at half-filling with $U \, / \, t = 8$ and different numbers of baths. SDR fit performed with a linear imaginary frequency grid with frequency cutoff $i\omega_{max} = 40\ t$ for $N_b = 8, 12, 24$ and $i\omega_{max} = 2\ t$ for $N_b = 16$.}
\label{fig:akw_2x2_adaptFreq}
\end{figure*}

Here, the calculation with 24 baths ($N_p = 6$) shows that the SDR fit fulfills this expectation. The spectral weights for the 24 bath calculation shown in Fig.~\ref{fig:akw_2x2_oldFreq} (lower right panel) coincide with the 8 and 12 bath calculations, and with the 16 bath calculations with accurately fitted low frequency behavior (see right panel of Fig.~\ref{fig:akw_2x2_diffGrids} and Fig.~\ref{fig:akw_2x2_adaptFreq} for a comparison of the spectral weights for all bath size with the 16 bath calculation using the shorter frequency grid). This finding suggests that the 24 bath calculation has fit the low frequency behavior of the hybridization function to the $10^{-4}$ accuracy mentioned above. Moreover, the red curve in the inset of Fig.~\ref{fig:diag_hyb_diffFreqs} shows the high frequency behavior of the hybridization function as computed with the converged bath parameters of the 24 bath calculation. Thus, it turns out that with 24 baths at its disposal, the SDR fit is capable of fitting both the high and low frequency behaviors, recovering the right physical behavior without any need for contrived constraints on the frequency grid in the fit. We have also performed a 24 bath calculation with the logarithmic frequency scale and the Matsubara scale ending at $i\omega = 2\, t$. The results did not change, adding further support to our claim.

As an additional note, it is also possible to add more weights to the low frequency range along the imaginary frequency axis. This can be done by adding a frequency dependent weight function $g(\omega)$ to the cost function in Eq.~\eqref{eq:CostFunc}. The effect of such a cost function is equivalent to choosing frequency grids of inhomogeneous densities, and therefore their effect does not qualitatively change the conclusions presented above.

We now summarize the main points of this subsection. The physical disadvantages of fitting on the imaginary frequency axis with a finite number of baths raise the important and nontrivial issue of deciding the frequency range for fitting. This choice may influence whether the results of the DMFT calculation are reliable and representative of the true physics of the system. We have found that one way to reach said physical  results without relying on \emph{a priori} knowledge to pick the best frequency range is to have a large number of baths and a corresponding fit that is able to treat such large bath effectively and in an unbiased way. We have shown that the SDR fit fulfills these conditions with a moderate number of poles ($N_p = 6$), and that paired with state-of-the art impurity solvers which can handle the large bath limit (ASCI in this paper), it can reach the physical set of bath parameters without choosing a tailored frequency range.

\section{Conclusion}
\label{sec:conclusion}

We have introduced an optimization method using semi-definite relaxation for the hybridization fitting step in the implementation of cluster DMFT. This optimization routine offers a straightforward approach to extrapolations into the infinite bath limit in Hamiltonian-based DMFT calculations. There has been development of other approaches with the same objective, notably an algorithm working directly on the real frequency axis~\cite{Lu2014} for single-site DMFT calculations, combining orbital rotations in the space of bath degrees of freedom with the substitution of the fitting step by a purely linear-algebraic calculation to treat hundreds of baths. There have also been advances toward employing compact Green's function representations on the imaginary frequency axis~\cite{Nagai2018} to extrapolate to the infinite bath limit in pure exact diagonalization DMFT. 

Our method is conceptually simpler than these approaches, mainly implementing an optimization routine to treat the most general bath structure effectively.  We have shown that it is an efficient and systematically improvable method, outperforming standard optimization routines and able to deal with a large number of bath parameters effectively. Furthermore, it has several important properties motivating its use in Hamiltonian-based cluster DMFT calculations: empirical robustness with respect to the initial guess, rapid convergence of low energy poles, apparent recognition of system symmetries, independence of the fitting frequency grid in the large bath limit, and versatility in the treatment of clusters with low symmetry. This method offers a systematic, natural, and easy-to-use approach to the fitting problem in DMFT, which despite its difficulty is perhaps the least-documented step in the self-consistency cycle.

\section*{Acknowledgments:} 

This work was partially supported by the Department of Energy under Grant No. DE-SC0017867, No. DE-AC02-05CH11231, by the Air Force Office of Scientific Research under award number FA9550-18-1-0095 (L.L. and L.Z.), by the National Science Foundation Graduate Research Fellowship Program under grant DGE-1106400 (M.L.), and by a \emph{Obra Social ``La Caixa''} graduate fellowship (ID
100010434), with code LCF/BQ/AA16/11580047 (C.M.Z.). Computational  resources provided  by  the  Extreme  Science  and  Engineering  Discovery Environment (XSEDE), which is supported by the National Science Foundation Grant No. OCI-1053575, are gratefully acknowledged. We thank Garnet Chan, Anil Damle, Alexandre Foley, Antoine Georges, Gabriel Kotliar, Olivier Parcollet, Reinhold Schneider, Lan Tran, Dominika Zgid and Jianxin Zhu for helpful discussions.

\newpage

\onecolumngrid

\setcounter{table}{0}
\renewcommand{\thetable}{S\arabic{table}}
\setcounter{figure}{0}
\renewcommand{\thefigure}{S\arabic{figure}}
\setcounter{equation}{0}
\renewcommand{\theequation}{S\arabic{equation}}

\section*{Supporting Information}

\subsection*{Bath parameters for the results in Sec. 3.2}

\begin{figure}[b]
\includegraphics{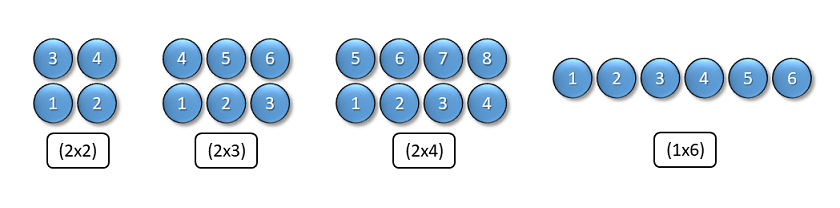}
\caption{Numbering convention for the cluster sites in the different clusters studied in this work. The numbers correspond to the $\alpha$ index in the $V_{\ell,\alpha}$ coupling terms.}
\label{fig:number_convention}
\end{figure}

Here we report the converged DMFT parameters obtained with the SDR fit and used to produce the data in section 3.2 of the main paper. These include fits for $(2\times2)$, $(2\times3)$, $(2\times4)$ and $(1\times6)$ cluster DMFT simulations. All are performed on a 2d square lattice Hubbard model at half-filling and $U\ / t = 8$. The impurity Hamiltonian follows
\begin{equation} \label{eq:Himp}
	H_{imp} = -t\sum_{\langle \alpha,\beta \rangle, \sigma} c^\dagger_{\alpha,\sigma} c_{\beta, \sigma} -\mu \sum_\alpha n_\alpha + U \sum_{\alpha} n_{\alpha, \uparrow} n_{\alpha, \downarrow} + \sum_{\ell=1}^{N_b}\epsilon_\ell\  d^\dagger_{\ell}d_{\ell} + \sum_{\ell=1}^{N_b}\sum_{\alpha=1}^{N_c}\left(V_{\alpha,\ell}\ d^\dagger_{\ell}c_{\alpha} + \mathrm{h.c.}\right)\ ,
\end{equation}
where $t$ is the nearest-neighbor hopping, $\mu$ is the chemical potential, $\epsilon_\ell$ are the bath energies and $V_{\ell,\alpha}$ the bath-cluster couplings. At half-filling $\mu = U / 2$. The hopping amplitude $t$ serves as our energy unit throughout.  Most of the fits were performed on a linear frequency grid of 50 points in the range $i\omega_n \in [0, 40\ t]$. The only exception are the $N_p = 4$ cases, where a grid of Matsubara frequencies with $\beta = 600$ and cut-off frequency of $i\omega_c = 2\ t$ was used. See main text for details. For the (2$\times$2) cluster and $N_p = 4$ we provide converged bath parameters for both grids, as well as a Matsubara grid with $\beta = 600$ and $i\omega_c = 10\ t$ and a logarithmic grid of 50 points in the interval $i\omega_n \in [0.01\ t, 40\ t]$. These were used to make Fig. 10 and 11 in the main paper. The numbering convention for the cluster sites is given in Fig.~\ref{fig:number_convention}.

\begin{table}[h!]
  \begin{center}
    \begin{tabular}{|r||c|c|c|c|} 
    \hline
      \textbf{$\epsilon_\ell$} & \textbf{$V_{\ell,1}$} & \textbf{$V_{\ell,2}$} & \textbf{$V_{\ell,3}$} & \textbf{$V_{\ell,4}$} \\
      \hline\hline
       	-3.2237 & -0.4009 &  0.4009 &  0.4009 & -0.4009\\ 
       	-3.2237 &  0.4751 &  0.4894 & -0.4894 & -0.4751\\ 
       	-3.2237 &  0.4894 & -0.4751 &  0.4751 & -0.4894\\ 
       	-3.2237 &  0.5525 &  0.5525 &  0.5525 &  0.5525\\
       	\hline\hline
       	 3.2237 & -0.4009 & -0.4009 & -0.4009 & -0.4009\\ 
       	 3.2237 & -0.4868 &  0.4778 & -0.4778 &  0.4868\\ 
       	 3.2237 &  0.4778 &  0.4868 & -0.4868 & -0.4778\\ 
       	 3.2237 &  0.5525 & -0.5525 & -0.5525 &  0.5525\\ 
		\hline
    \end{tabular}
  \end{center}
  \caption{Bath parameters for a cluster DMFT calculation on the 2d square lattice Hubbard model at half-filling and $U\ / \ t = 8$ for a (2$\times$2) cluster with $N_p = 2$. These were obtained with a linear frequency grid of 50 points in the interval $i\omega_n \in [0, 40\ t]$.}
  \label{Table:2x2_Np2}
\end{table}

\begin{table}[h!]
    \begin{center}
        \begin{tabular}{|r||c|c|c|c|}
        \hline
          \textbf{$\epsilon_\ell$} & \textbf{$V_{\ell,1}$} & \textbf{$V_{\ell,2}$} & \textbf{$V_{\ell,3}$} & \textbf{$V_{\ell,4}$} \\
          \hline\hline
            -3.3009 & -0.3983 &  0.3983 &  0.3983 & -0.3983 \\
            -3.3009 &  0.6773 &  0.0691 & -0.0691 & -0.6773 \\
            -3.3009 & -0.0691 &  0.6773 & -0.6773 &  0.0691 \\
            -3.3009 &  0.5526 &  0.5527 &  0.5527 &  0.5526 \\
            \hline\hline
            -0.0000 &  0.0790 &  0.0538 & -0.0536 & -0.0790 \\
            -0.0000 &  0.0537 & -0.0790 &  0.0791 & -0.0537 \\
            -0.0000 & -0.0963 &  0.0095 &  0.0095 & -0.0963 \\
            -0.0000 & -0.0094 & -0.0963 & -0.0963 & -0.0096 \\
            \hline\hline
             3.3009 & -0.3983 & -0.3983 & -0.3983 & -0.3983 \\
             3.3009 &  0.0723 & -0.6770 &  0.6770 & -0.0723 \\
             3.3009 &  0.6770 &  0.0723 & -0.0723 & -0.6770 \\
             3.3009 &  0.5527 & -0.5526 & -0.5526 &  0.5527 \\
            \hline
        \end{tabular}
    \end{center}
    \caption{Bath parameters for a cluster DMFT calculation on the 2d square lattice Hubbard model at half-filling and $U\ / \ t = 8$ for a (2$\times$2) cluster with $N_p = 3$. These were obtained with a linear frequency grid of 50 points in the interval $i\omega_n \in [0, 40\ t]$.}
    \label{Table:2x2_Np3}
\end{table}

\begin{table}[h!]
    \begin{center}
        \begin{tabular}{|r||c|c|c|c|}
        \hline
          \textbf{$\epsilon_\ell$} & \textbf{$V_{\ell,1}$} & \textbf{$V_{\ell,2}$} & \textbf{$V_{\ell,3}$} & \textbf{$V_{\ell,4}$} \\
          \hline\hline
            -3.3418 & -0.3142 &  0.3142 &  0.3142 & -0.3142 \\
            -3.3418 &  0.0148 &  0.5700 & -0.5700 & -0.0148 \\
            -3.3418 &  0.5700 & -0.0148 &  0.0148 & -0.5700 \\
            -3.3418 &  0.4761 &  0.4761 &  0.4761 &  0.4761 \\
            \hline\hline
            -1.5618 & -0.1790 &  0.1791 &  0.1790 & -0.1790 \\
            -1.5618 &  0.2618 &  0.0864 & -0.0864 & -0.2618 \\
            -1.5618 & -0.0864 &  0.2618 & -0.2618 &  0.0864 \\
            -1.5618 &  0.2123 &  0.2123 &  0.2123 &  0.2123 \\
            \hline\hline
             1.5618 & -0.1790 & -0.1790 & -0.1790 & -0.1790 \\
             1.5618 &  0.2539 & -0.1075 &  0.1075 & -0.2539 \\
             1.5618 & -0.1075 & -0.2539 &  0.2539 &  0.1075 \\
             1.5618 &  0.2123 & -0.2123 & -0.2123 &  0.2123 \\
            \hline\hline
             3.3418 &  0.3142 &  0.3142 &  0.3142 &  0.3142 \\
             3.3418 & -0.1675 & -0.5450 &  0.5450 &  0.1675 \\
             3.3418 & -0.5450 &  0.1675 & -0.1675 &  0.5450 \\
             3.3418 &  0.4761 & -0.4761 & -0.4761 &  0.4761 \\
            \hline
        \end{tabular}
    \end{center}
    \caption{Bath parameters for a cluster DMFT calculation on the 2d square lattice Hubbard model at half-filling and $U\ / \ t = 8$ for a (2$\times$2) cluster with $N_p = 4$. These were obtained with a Matsubara frequency grid with $\beta = 600$ and cut-off frequency $i\omega_c = 2\ t$.}
    \label{Table:2x2_Np4_mats2}
\end{table}

\begin{table}[h!]
    \begin{center}
        \begin{tabular}{|r||c|c|c|c|}
        \hline
          \textbf{$\epsilon_\ell$} & \textbf{$V_{\ell,1}$} & \textbf{$V_{\ell,2}$} & \textbf{$V_{\ell,3}$} & \textbf{$V_{\ell,4}$} \\
          \hline\hline
            -5.0614 &  0.3333 & -0.3333 & -0.3332 &  0.3333 \\
            -5.0614 &  0.4480 & -0.3433 &  0.3433 & -0.4480 \\
            -5.0614 & -0.3433 & -0.4480 &  0.4480 &  0.3433 \\
            -5.0614 &  0.4529 &  0.4529 &  0.4529 &  0.4529 \\
            \hline\hline
            -1.9679 & -0.2477 &  0.2477 &  0.2477 & -0.2477 \\
            -1.9679 &  0.2693 &  0.3249 & -0.3249 & -0.2693 \\
            -1.9679 &  0.3249 & -0.2693 &  0.2693 & -0.3249 \\
            -1.9679 &  0.3446 &  0.3446 &  0.3446 &  0.3446 \\
            \hline\hline
             1.9685 & -0.2478 & -0.2478 & -0.2478 & -0.2478 \\
             1.9685 & -0.3452 &  0.2430 & -0.2430 &  0.3452 \\
             1.9685 &  0.2430 &  0.3452 & -0.3452 & -0.2430 \\
             1.9685 &  0.3447 & -0.3447 & -0.3447 &  0.3447 \\
            \hline\hline
             5.0623 &  0.3332 &  0.3332 &  0.3332 &  0.3332 \\
             5.0623 & -0.2390 & -0.5112 &  0.5112 &  0.2390 \\
             5.0623 & -0.5112 &  0.2390 & -0.2390 &  0.5112 \\
             5.0623 &  0.4529 & -0.4529 & -0.4529 &  0.4529 \\
            \hline
        \end{tabular}
    \end{center}
    \caption{Bath parameters for a cluster DMFT calculation on the 2d square lattice Hubbard model at half-filling and $U\ / \ t = 8$ for a (2$\times$2) cluster with $N_p = 4$. These were obtained with a linear frequency grid of 50 points in the interval $i\omega_n \in [0, 40\ t]$.}
    \label{Table:2x2_Np4_lin}
\end{table}

\begin{table}[h!]
    \begin{center}
        \begin{tabular}{|r||c|c|c|c|}
        \hline
          \textbf{$\epsilon_\ell$} & \textbf{$V_{\ell,1}$} & \textbf{$V_{\ell,2}$} & \textbf{$V_{\ell,3}$} & \textbf{$V_{\ell,4}$} \\
          \hline\hline
            -4.7209 & -0.3356 &  0.3356 &  0.3356 & -0.3356 \\
            -4.7209 & -0.4121 &  0.4124 & -0.4124 &  0.4121 \\
            -4.7209 & -0.4124 & -0.4121 &  0.4121 &  0.4124 \\
            -4.7209 &  0.4742 &  0.4742 &  0.4742 &  0.4742 \\
            \hline\hline
            -1.8420 & -0.2350 &  0.2350 &  0.2350 & -0.2350 \\
            -1.8420 & -0.2757 & -0.2736 &  0.2736 &  0.2757 \\
            -1.8420 &  0.2736 & -0.2757 &  0.2757 & -0.2736 \\
            -1.8420 &  0.3125 &  0.3125 &  0.3125 &  0.3125 \\
            \hline\hline
             1.8420 & -0.2350 & -0.2350 & -0.2350 & -0.2350 \\
             1.8420 &  0.2699 & -0.2793 &  0.2793 & -0.2699 \\
             1.8420 &  0.2793 &  0.2699 & -0.2699 & -0.2793 \\
             1.8420 &  0.3125 & -0.3125 & -0.3125 &  0.3125 \\
            \hline\hline
             4.7209 &  0.3356 &  0.3356 &  0.3356 &  0.3356 \\
             4.7209 & -0.4276 & -0.3963 &  0.3963 &  0.4276 \\
             4.7209 &  0.3963 & -0.4276 &  0.4276 & -0.3963 \\
             4.7209 &  0.4742 & -0.4742 & -0.4742 &  0.4742 \\
            \hline
        \end{tabular}
    \end{center}
    \caption{Bath parameters for a cluster DMFT calculation on the 2d square lattice Hubbard model at half-filling and $U\ / \ t = 8$ for a (2$\times$2) cluster with $N_p = 4$. These were obtained with a logarithmic frequency grid of 50 points in the interval $i\omega_n \in [0.01\ t, 40\ t]$.}
    \label{Table:2x2_Np4_log}
\end{table}

\begin{table}[h!]
    \begin{center}
        \begin{tabular}{|r||c|c|c|c|}
        \hline
          \textbf{$\epsilon_\ell$} & \textbf{$V_{\ell,1}$} & \textbf{$V_{\ell,2}$} & \textbf{$V_{\ell,3}$} & \textbf{$V_{\ell,4}$} \\
          \hline\hline
            -4.5944 & -0.3399 &  0.3399 &  0.3399 & -0.3399 \\
            -4.5944 & -0.3998 &  0.4244 & -0.4244 &  0.3998 \\
            -4.5944 & -0.4244 & -0.3998 &  0.3998 &  0.4244 \\
            -4.5944 &  0.4711 &  0.4711 &  0.4711 &  0.4711 \\
            \hline\hline
            -1.8332 & -0.2266 &  0.2266 &  0.2266 & -0.2266 \\
            -1.8332 & -0.2728 & -0.2663 &  0.2663 &  0.2728 \\
            -1.8332 &  0.2663 & -0.2728 &  0.2728 & -0.2663 \\
            -1.8332 &  0.3099 &  0.3099 &  0.3099 &  0.3099 \\
            \hline\hline
             1.8332 & -0.2266 & -0.2266 & -0.2266 & -0.2266 \\
             1.8332 & -0.2711 &  0.2679 & -0.2679 &  0.2711 \\
             1.8332 &  0.2679 &  0.2711 & -0.2711 & -0.2679 \\
             1.8332 &  0.3099 & -0.3099 & -0.3099 &  0.3099 \\
            \hline\hline
             4.5944 &  0.3399 &  0.3399 &  0.3399 &  0.3399 \\
             4.5944 & -0.4002 & -0.4240 &  0.4240 &  0.4002 \\
             4.5944 & -0.4240 &  0.4002 & -0.4002 &  0.4240 \\
             4.5944 &  0.4711 & -0.4711 & -0.4711 &  0.4711 \\
            \hline
        \end{tabular}
    \end{center}
    \caption{Bath parameters for a cluster DMFT calculation on the 2d square lattice Hubbard model at half-filling and $U\ / \ t = 8$ for a (2$\times$2) cluster with $N_p = 4$. These were obtained with a Matsubara frequency grid with $\beta = 600$ and cut-off frequency $i\omega_c = 10\ t$.}
    \label{Table:2x2_Np4_mats10}
\end{table}

\begin{table}[h!]
    \begin{center}
        \begin{tabular}{|r||c|c|c|c|}
        \hline
          \textbf{$\epsilon_\ell$} & \textbf{$V_{\ell,1}$} & \textbf{$V_{\ell,2}$} & \textbf{$V_{\ell,3}$} & \textbf{$V_{\ell,4}$} \\
          \hline\hline
            -6.4988 &  0.2714 & -0.2714 & -0.2714 &  0.2714 \\
            -6.4988 &  0.2971 &  0.2609 & -0.2609 & -0.2971 \\
            -6.4988 & -0.2609 &  0.2971 & -0.2972 &  0.2609 \\
            -6.4988 &  0.2872 &  0.2872 &  0.2872 &  0.2872 \\
            \hline\hline
            -3.5644 & -0.2405 &  0.2405 &  0.2405 & -0.2405 \\
            -3.5644 & -0.3278 &  0.3563 & -0.3563 &  0.3278 \\
            -3.5644 & -0.3563 & -0.3278 &  0.3278 &  0.3563 \\
            -3.5644 &  0.4178 &  0.4178 &  0.4178 &  0.4178 \\
            \hline\hline
            -1.6967 & -0.2120 &  0.2120 &  0.2120 & -0.2120 \\
            -1.6967 & -0.2383 & -0.2290 &  0.2290 &  0.2383 \\
            -1.6967 &  0.2290 & -0.2383 &  0.2383 & -0.2290 \\
            -1.6967 &  0.2576 &  0.2576 &  0.2576 &  0.2576 \\
            \hline\hline
             1.6972 & -0.2120 & -0.2120 & -0.2121 & -0.2120 \\
             1.6972 &  0.2141 & -0.2519 &  0.2519 & -0.2141 \\
             1.6972 &  0.2519 &  0.2141 & -0.2141 & -0.2519 \\
             1.6972 &  0.2577 & -0.2577 & -0.2577 &  0.2577 \\
            \hline\hline
             3.5650 &  0.2404 &  0.2404 &  0.2404 &  0.2404 \\
             3.5650 & -0.4604 & -0.1495 &  0.1495 &  0.4604 \\
             3.5650 &  0.1495 & -0.4604 &  0.4604 & -0.1495 \\
             3.5650 &  0.4177 & -0.4177 & -0.4177 &  0.4177 \\
            \hline\hline
             6.4987 &  0.2714 &  0.2714 &  0.2714 &  0.2714 \\
             6.4987 & -0.0384 & -0.3935 &  0.3935 &  0.0384 \\
             6.4987 & -0.3935 &  0.0384 & -0.0384 &  0.3935 \\
             6.4987 &  0.2872 & -0.2872 & -0.2872 &  0.2872 \\
            \hline
        \end{tabular}
    \end{center}
    \caption{Bath parameters for a cluster DMFT calculation on the 2d square lattice Hubbard model at half-filling and $U\ / \ t = 8$ for a (2$\times$2) cluster with $N_p = 6$. These were obtained with a linear frequency grid of 50 points in the interval $i\omega_n \in [0, 40\ t]$.}
    \label{Table:2x2_Np6}
\end{table}

\begin{table}[h!]
    \begin{center}
        \begin{tabular}{|r||c|c|c|c|c|c|}
        \hline
          \textbf{$\epsilon_\ell$} & \textbf{$V_{\ell,1}$} & \textbf{$V_{\ell,2}$} & \textbf{$V_{\ell,3}$} & \textbf{$V_{\ell,4}$} & \textbf{$V_{\ell,5}$} & \textbf{$V_{\ell,6}$} \\
          \hline\hline
            -3.1463 &  0.0477 & -0.1391 &  0.4342 &  0.4342 & -0.1391 &  0.0477 \\
            -3.1463 & -0.0370 &  0.1390 &  0.4366 & -0.4366 & -0.1390 &  0.0370 \\
            -3.1463 & -0.4657 &  0.3830 & -0.1613 &  0.1613 & -0.3830 &  0.4657 \\
            -3.1463 &  0.4776 & -0.3823 & -0.1750 & -0.1750 & -0.3823 &  0.4776 \\
            -3.1463 &  0.4808 &  0.5449 &  0.1218 &  0.1218 &  0.5449 &  0.4808 \\
            -3.1463 & -0.4940 & -0.5451 &  0.1316 & -0.1316 &  0.5451 &  0.4940 \\
            \hline\hline
             3.1515 & -0.0611 & -0.1431 & -0.4301 & -0.4301 & -0.1431 & -0.0611 \\
             3.1515 & -0.0279 & -0.1351 &  0.4399 & -0.4399 &  0.1351 &  0.0279 \\
             3.1515 & -0.4620 & -0.3802 &  0.1921 &  0.1921 & -0.3802 & -0.4620 \\
             3.1515 & -0.4765 & -0.3883 & -0.1495 &  0.1495 &  0.3883 &  0.4765 \\
             3.1515 & -0.4837 &  0.5413 &  0.1357 & -0.1357 & -0.5413 &  0.4837 \\
             3.1515 &  0.4953 & -0.5458 &  0.1112 &  0.1112 & -0.5458 &  0.4953 \\
            \hline
        \end{tabular}
    \end{center}
    \caption{Bath parameters for a cluster DMFT calculation on the 2d square lattice Hubbard model at half-filling and $U\ / \ t = 8$ for a (2$\times$3) cluster with $N_p = 2$. These were obtained with a linear frequency grid of 50 points in the interval $i\omega_n \in [0, 40\ t]$.}
    \label{Table:2x3_Np2}
\end{table}

\begin{table}[h!]
    \begin{center}
        \begin{tabular}{|r||c|c|c|c|c|c|}
        \hline
          \textbf{$\epsilon_\ell$} & \textbf{$V_{\ell,1}$} & \textbf{$V_{\ell,2}$} & \textbf{$V_{\ell,3}$} & \textbf{$V_{\ell,4}$} & \textbf{$V_{\ell,5}$} & \textbf{$V_{\ell,6}$} \\
          \hline\hline
            -3.3942 &  0.0604 & -0.1295 & -0.3564 &  0.3564 &  0.1295 & -0.0604 \\
            -3.3942 &  0.0377 & -0.1328 &  0.3607 &  0.3607 & -0.1328 &  0.0377 \\
            -3.3942 &  0.3824 & -0.3023 & -0.1512 & -0.1512 & -0.3023 &  0.3824 \\
            -3.3942 &  0.4018 & -0.3012 &  0.1776 & -0.1776 &  0.3012 & -0.4018 \\
            -3.3942 &  0.3876 &  0.4577 & -0.1006 &  0.1006 & -0.4577 & -0.3876 \\
            -3.3942 &  0.4117 &  0.4580 &  0.1255 &  0.1255 &  0.4580 &  0.4117 \\
            \hline\hline
            -1.6540 &  0.0393 & -0.0526 &  0.1822 &  0.1822 & -0.0526 &  0.0393 \\
            -1.6540 &  0.0263 &  0.0394 &  0.1911 & -0.1911 & -0.0394 & -0.0263 \\
            -1.6540 & -0.1794 &  0.1720 & -0.0107 &  0.0107 & -0.1720 &  0.1794 \\
            -1.6540 & -0.2531 &  0.1319 &  0.0927 &  0.0927 &  0.1319 & -0.2531 \\
            -1.6540 & -0.1534 & -0.2640 & -0.0431 & -0.0431 & -0.2640 & -0.1534 \\
            -1.6540 & -0.2413 & -0.2465 &  0.0840 & -0.0840 &  0.2465 &  0.2413 \\
            \hline\hline
             1.6943 & -0.0616 & -0.0659 & -0.1737 & -0.1737 & -0.0659 & -0.0616 \\
             1.6943 &  0.0272 & -0.0354 &  0.2017 & -0.2017 &  0.0354 & -0.0272 \\
             1.6943 & -0.1697 & -0.1684 &  0.1241 &  0.1241 & -0.1684 & -0.1697 \\
             1.6943 & -0.2074 & -0.2112 & -0.0091 &  0.0091 &  0.2112 &  0.2074 \\
             1.6943 & -0.2280 &  0.2210 &  0.0696 & -0.0696 & -0.2210 &  0.2280 \\
             1.6943 &  0.2578 & -0.2557 &  0.0055 &  0.0055 & -0.2557 &  0.2578 \\
            \hline\hline
             3.4497 &  0.0607 &  0.1445 & -0.3420 &  0.3420 & -0.1445 & -0.0607 \\
             3.4497 & -0.0307 & -0.1374 & -0.3538 & -0.3538 & -0.1374 & -0.0307 \\
             3.4497 &  0.3911 &  0.2811 &  0.1882 & -0.1882 & -0.2811 & -0.3911 \\
             3.4497 &  0.4203 &  0.2821 & -0.1461 & -0.1461 &  0.2821 &  0.4203 \\
             3.4497 &  0.3609 & -0.4611 &  0.1478 &  0.1478 & -0.4611 &  0.3609 \\
             3.4497 & -0.3947 &  0.4646 &  0.1261 & -0.1261 & -0.4646 &  0.3947 \\
            \hline
        \end{tabular}
    \end{center}
    \caption{Bath parameters for a cluster DMFT calculation on the 2d square lattice Hubbard model at half-filling and $U\ / \ t = 8$ for a (2$\times$3) cluster with $N_p = 4$. These were obtained with a Matsubara frequency grid with $\beta = 600$ and cut-off frequency $i\omega_c = 2\ t$.}
    \label{Table:2x3_Np4}
\end{table}

\begin{table}[h!]
    \begin{center}
        \begin{tabular}{|r||c|c|c|c|c|c|}
        \hline
          \textbf{$\epsilon_\ell$} & \textbf{$V_{\ell,1}$} & \textbf{$V_{\ell,2}$} & \textbf{$V_{\ell,3}$} & \textbf{$V_{\ell,4}$} & \textbf{$V_{\ell,5}$} & \textbf{$V_{\ell,6}$} \\
          \hline\hline
            -6.6130 & -0.0074 &  0.0250 &  0.2519 & -0.2519 & -0.0250 &  0.0074 \\
            -6.6130 & -0.0098 & -0.0214 &  0.2524 &  0.2524 & -0.0214 & -0.0098 \\
            -6.6130 & -0.2623 &  0.2680 &  0.0125 &  0.0125 &  0.2680 & -0.2623 \\
            -6.6130 & -0.2602 &  0.2880 & -0.0361 &  0.0361 & -0.2880 &  0.2602 \\
            -6.6130 & -0.2921 & -0.2617 &  0.0174 & -0.0174 &  0.2617 &  0.2921 \\
            -6.6130 &  0.2870 &  0.2793 &  0.0349 &  0.0349 &  0.2793 &  0.2870 \\
            \hline\hline
            -3.6121 &  0.0711 & -0.1254 & -0.2763 &  0.2763 &  0.1254 & -0.0711 \\
            -3.6121 &  0.0438 & -0.1309 &  0.2859 &  0.2859 & -0.1309 &  0.0438 \\
            -3.6121 &  0.3104 & -0.2336 & -0.1546 & -0.1546 & -0.2336 &  0.3104 \\
            -3.6121 &  0.3278 & -0.2337 &  0.1904 & -0.1904 &  0.2337 & -0.3278 \\
            -3.6121 &  0.3265 &  0.3842 & -0.0905 &  0.0905 & -0.3842 & -0.3265 \\
            -3.6121 &  0.3527 &  0.3871 &  0.1232 &  0.1232 &  0.3871 &  0.3527 \\
            \hline\hline
            -1.8081 & -0.0443 &  0.0658 & -0.2225 & -0.2225 &  0.0658 & -0.0443 \\
            -1.8081 & -0.0223 & -0.0535 & -0.2330 &  0.2330 &  0.0535 &  0.0223 \\
            -1.8081 & -0.2231 &  0.2047 & -0.0256 &  0.0256 & -0.2047 &  0.2231 \\
            -1.8081 & -0.2841 &  0.1772 &  0.1090 &  0.1090 &  0.1772 & -0.2841 \\
            -1.8081 & -0.2081 & -0.3037 & -0.0485 & -0.0485 & -0.3037 & -0.2081 \\
            -1.8081 & -0.2792 & -0.2924 &  0.0938 & -0.0938 &  0.2924 &  0.2792 \\
            \hline\hline
             1.8180 & -0.0649 & -0.0738 & -0.2085 & -0.2085 & -0.0738 & -0.0649 \\
             1.8180 &  0.0252 & -0.0448 &  0.2344 & -0.2344 &  0.0448 & -0.0252 \\
             1.8180 & -0.2043 & -0.1952 &  0.1326 &  0.1326 & -0.1952 & -0.2043 \\
             1.8180 & -0.2466 & -0.2299 & -0.0174 &  0.0174 &  0.2299 &  0.2466 \\
             1.8180 & -0.2516 &  0.2640 &  0.0775 & -0.0775 & -0.2640 &  0.2516 \\
             1.8180 &  0.2863 & -0.2903 &  0.0137 &  0.0137 & -0.2903 &  0.2863 \\
            \hline\hline
             3.5287 &  0.0614 &  0.1393 & -0.2607 &  0.2607 & -0.1393 & -0.0614 \\
             3.5287 & -0.0446 & -0.1356 & -0.2696 & -0.2696 & -0.1356 & -0.0446 \\
             3.5287 &  0.3187 &  0.2045 &  0.1843 & -0.1843 & -0.2045 & -0.3187 \\
             3.5287 &  0.3490 &  0.1994 & -0.1580 & -0.1580 &  0.1994 &  0.3490 \\
             3.5287 &  0.2881 & -0.3875 &  0.1473 &  0.1473 & -0.3875 &  0.2881 \\
             3.5287 & -0.3269 &  0.3907 &  0.1318 & -0.1318 & -0.3907 &  0.3269 \\
            \hline\hline
             6.4752 &  0.0284 &  0.0219 & -0.2572 &  0.2572 & -0.0219 & -0.0284 \\
             6.4752 &  0.0239 & -0.0175 & -0.2585 & -0.2585 & -0.0175 &  0.0239 \\
             6.4752 &  0.2580 &  0.2836 &  0.0527 & -0.0527 & -0.2836 & -0.2580 \\
             6.4752 & -0.2486 & -0.3142 & -0.0018 & -0.0018 & -0.3142 & -0.2486 \\
             6.4752 & -0.3214 &  0.2545 & -0.0469 & -0.0469 &  0.2545 & -0.3214 \\
             6.4752 & -0.3090 &  0.2830 & -0.0101 &  0.0101 & -0.2830 &  0.3090 \\
            \hline
        \end{tabular}
    \end{center}
    \caption{Bath parameters for a cluster DMFT calculation on the 2d square lattice Hubbard model at half-filling and $U\ / \ t = 8$ for a (2$\times$3) cluster with $N_p = 6$. These were obtained with a linear frequency grid of 50 points in the interval $i\omega_n \in [0, 40\ t]$.}
    \label{Table:2x3_Np6}
\end{table}

\begin{table}[h!]
    \begin{center}
        \begin{tabular}{|r||c|c|c|c|c|c|c|c|}
        \hline
          \textbf{$\epsilon_\ell$} & \textbf{$V_{\ell,1}$} & \textbf{$V_{\ell,2}$} & \textbf{$V_{\ell,3}$} & \textbf{$V_{\ell,4}$} & \textbf{$V_{\ell,5}$} & \textbf{$V_{\ell,6}$} & \textbf{$V_{\ell,7}$} & \textbf{$V_{\ell,8}$} \\
          \hline\hline
            -3.1946 & -0.1226 &  0.1212 &  0.2818 & -0.3135 & -0.3135 &  0.2818 &  0.1212 & -0.1226 \\
            -3.1946 &  0.1031 & -0.0985 & -0.3383 &  0.2853 & -0.2853 &  0.3383 &  0.0985 & -0.1031 \\
            -3.1946 &  0.0696 &  0.0727 & -0.2959 & -0.3509 &  0.3509 &  0.2959 & -0.0727 & -0.0696 \\
            -3.1946 & -0.0543 & -0.0539 &  0.3475 &  0.3128 &  0.3128 &  0.3475 & -0.0539 & -0.0543 \\
            -3.1946 &  0.4224 & -0.4248 &  0.1454 & -0.1269 &  0.1269 & -0.1454 &  0.4248 & -0.4224 \\
            -3.1946 &  0.4190 & -0.4214 &  0.1626 & -0.1806 & -0.1806 &  0.1626 & -0.4214 &  0.4190 \\
            -3.1946 &  0.5209 &  0.5193 &  0.0910 &  0.0788 &  0.0788 &  0.0910 &  0.5193 &  0.5209 \\
            -3.1946 & -0.5206 & -0.5191 & -0.1083 & -0.1194 &  0.1194 &  0.1083 &  0.5191 &  0.5206 \\
            \hline\hline
             3.1946 & -0.1226 & -0.1212 & -0.2817 & -0.3135 & -0.3135 & -0.2817 & -0.1212 & -0.1226 \\
             3.1946 &  0.1031 &  0.0985 &  0.3381 &  0.2855 & -0.2855 & -0.3381 & -0.0985 & -0.1031 \\
             3.1946 &  0.0696 & -0.0727 &  0.2961 & -0.3508 &  0.3508 & -0.2961 &  0.0727 & -0.0696 \\
             3.1946 & -0.0543 &  0.0539 & -0.3476 &  0.3127 &  0.3127 & -0.3476 &  0.0539 & -0.0543 \\
             3.1946 &  0.4224 &  0.4248 & -0.1454 & -0.1269 &  0.1269 &  0.1454 & -0.4248 & -0.4224 \\
             3.1946 &  0.4190 &  0.4214 & -0.1626 & -0.1806 & -0.1806 & -0.1626 &  0.4214 &  0.4190 \\
             3.1946 &  0.5209 & -0.5194 & -0.0910 &  0.0788 &  0.0788 & -0.0910 & -0.5194 &  0.5209 \\
             3.1946 & -0.5206 &  0.5191 &  0.1083 & -0.1194 &  0.1194 & -0.1083 & -0.5191 &  0.5206 \\
            \hline
        \end{tabular}
    \end{center}
    \caption{Bath parameters for a cluster DMFT calculation on the 2d square lattice Hubbard model at half-filling and $U\ / \ t = 8$ for a (2$\times$4) cluster with $N_p = 2$. These were obtained with a linear frequency grid of 50 points in the interval $i\omega_n \in [0, 40\ t]$.}
    \label{Table:2x4_Np2}
\end{table}

\begin{table}[h!]
    \begin{center}
        \begin{tabular}{|r||c|c|c|c|c|c|c|c|}
        \hline
          \textbf{$\epsilon_\ell$} & \textbf{$V_{\ell,1}$} & \textbf{$V_{\ell,2}$} & \textbf{$V_{\ell,3}$} & \textbf{$V_{\ell,4}$} & \textbf{$V_{\ell,5}$} & \textbf{$V_{\ell,6}$} & \textbf{$V_{\ell,7}$} & \textbf{$V_{\ell,8}$} \\
          \hline\hline
            -3.4661 & -0.0753 &  0.0853 &  0.2243 & -0.2255 & -0.2255 &  0.2243 &  0.0853 & -0.0753 \\
            -3.4661 &  0.0322 &  0.0616 & -0.2349 & -0.2497 &  0.2497 &  0.2349 & -0.0616 & -0.0322 \\
            -3.4661 & -0.2180 &  0.2006 &  0.2150 & -0.1808 &  0.1808 & -0.2150 & -0.2006 &  0.2180 \\
            -3.4661 & -0.1744 & -0.1331 &  0.2768 &  0.2831 &  0.2831 &  0.2768 & -0.1331 & -0.1744 \\
            -3.4661 &  0.3329 & -0.3557 &  0.1438 & -0.1028 & -0.1028 &  0.1438 & -0.3557 &  0.3329 \\
            -3.4661 &  0.2581 & -0.2903 &  0.2793 & -0.3010 &  0.3010 & -0.2793 &  0.2903 & -0.2581 \\
            -3.4661 &  0.4291 &  0.4141 &  0.1012 &  0.0624 & -0.0624 & -0.1012 & -0.4141 & -0.4291 \\
            -3.4661 &  0.3888 &  0.3834 &  0.2051 &  0.2192 &  0.2192 &  0.2051 &  0.3834 &  0.3888 \\
            \hline\hline
            -1.9316 & -0.0465 &  0.0237 &  0.0525 & -0.1189 & -0.1189 &  0.0525 &  0.0237 & -0.0465 \\
            -1.9316 &  0.0465 &  0.0008 & -0.0376 & -0.1319 &  0.1319 &  0.0376 & -0.0008 & -0.0465 \\
            -1.9316 &  0.0162 & -0.0326 &  0.1643 & -0.0413 &  0.0413 & -0.1643 &  0.0326 & -0.0162 \\
            -1.9316 & -0.0133 & -0.0215 & -0.1604 & -0.0699 & -0.0699 & -0.1604 & -0.0215 & -0.0133 \\
            -1.9316 & -0.1787 &  0.2243 & -0.0546 &  0.0905 &  0.0905 & -0.0546 &  0.2243 & -0.1787 \\
            -1.9316 & -0.1815 &  0.2345 &  0.0454 & -0.0755 &  0.0755 & -0.0454 & -0.2345 &  0.1815 \\
            -1.9316 & -0.2731 & -0.2312 & -0.0407 & -0.0860 &  0.0860 &  0.0407 &  0.2312 &  0.2731 \\
            -1.9316 &  0.2761 &  0.2423 & -0.0247 & -0.0704 & -0.0704 & -0.0247 &  0.2423 &  0.2761 \\
            \hline\hline
             1.9318 & -0.0465 & -0.0237 & -0.0525 & -0.1189 & -0.1189 & -0.0525 & -0.0237 & -0.0465 \\
             1.9318 &  0.0465 & -0.0008 &  0.0377 & -0.1319 &  0.1319 & -0.0377 &  0.0008 & -0.0465 \\
             1.9318 &  0.0162 &  0.0326 & -0.1642 & -0.0414 &  0.0414 &  0.1642 & -0.0326 & -0.0162 \\
             1.9318 & -0.0134 &  0.0215 &  0.1604 & -0.0699 & -0.0699 &  0.1604 &  0.0215 & -0.0134 \\
             1.9318 & -0.1787 & -0.2244 &  0.0546 &  0.0905 &  0.0905 &  0.0546 & -0.2244 & -0.1787 \\
             1.9318 & -0.1814 & -0.2345 & -0.0454 & -0.0754 &  0.0754 &  0.0454 &  0.2345 &  0.1814 \\
             1.9318 & -0.2732 &  0.2312 &  0.0406 & -0.0860 &  0.0860 & -0.0406 & -0.2312 &  0.2732 \\
             1.9318 &  0.2762 & -0.2422 &  0.0247 & -0.0705 & -0.0705 &  0.0247 & -0.2422 &  0.2762 \\
            \hline\hline
             3.4663 &  0.0754 &  0.0853 &  0.2243 &  0.2255 &  0.2255 &  0.2243 &  0.0853 &  0.0754 \\
             3.4663 &  0.0322 & -0.0616 &  0.2349 & -0.2497 &  0.2497 & -0.2349 &  0.0616 & -0.0322 \\
             3.4663 &  0.2181 &  0.2006 &  0.2150 &  0.1808 & -0.1808 & -0.2150 & -0.2006 & -0.2181 \\
             3.4663 &  0.1745 & -0.1330 &  0.2767 & -0.2831 & -0.2831 &  0.2767 & -0.1330 &  0.1745 \\
             3.4663 & -0.3328 & -0.3557 &  0.1439 &  0.1027 &  0.1027 &  0.1439 & -0.3557 & -0.3328 \\
             3.4663 & -0.2581 & -0.2902 &  0.2793 &  0.3010 & -0.3010 & -0.2793 &  0.2902 &  0.2581 \\
             3.4663 & -0.4290 &  0.4142 &  0.1012 & -0.0624 &  0.0624 & -0.1012 & -0.4142 &  0.4290 \\
             3.4663 &  0.3888 & -0.3834 & -0.2052 &  0.2192 &  0.2192 & -0.2052 & -0.3834 &  0.3888 \\
            \hline
        \end{tabular}
    \end{center}
    \caption{Bath parameters for a cluster DMFT calculation on the 2d square lattice Hubbard model at half-filling and $U\ / \ t = 8$ for a (2$\times$4) cluster with $N_p = 4$. These were obtained with a Matsubara frequency grid with $\beta = 600$ and cut-off frequency $i\omega_c = 2\ t$.}
    \label{Table:2x4_Np4}
\end{table}

\begin{table}[h!]
    \begin{center}
        \begin{tabular}{|r||c|c|c|c|c|c|c|c|}
        \hline
          \textbf{$\epsilon_\ell$} & \textbf{$V_{\ell,1}$} & \textbf{$V_{\ell,2}$} & \textbf{$V_{\ell,3}$} & \textbf{$V_{\ell,4}$} & \textbf{$V_{\ell,5}$} & \textbf{$V_{\ell,6}$} & \textbf{$V_{\ell,7}$} & \textbf{$V_{\ell,8}$} \\
          \hline\hline
            -6.1207 & -0.0038 &  0.0083 & -0.1689 & -0.1769 &  0.1769 &  0.1689 & -0.0083 &  0.0038 \\
            -6.1207 & -0.0041 & -0.0056 & -0.1698 &  0.1762 &  0.1762 & -0.1698 & -0.0056 & -0.0041 \\
            -6.1207 & -0.0534 &  0.0376 &  0.2101 & -0.1977 &  0.1977 & -0.2101 & -0.0376 &  0.0534 \\
            -6.1207 & -0.0477 & -0.0269 &  0.2116 &  0.2020 &  0.2020 &  0.2116 & -0.0269 & -0.0477 \\
            -6.1207 & -0.2898 &  0.3098 & -0.0151 & -0.0114 & -0.0114 & -0.0151 &  0.3098 & -0.2898 \\
            -6.1207 &  0.2406 & -0.3590 &  0.0552 & -0.0747 &  0.0747 & -0.0552 &  0.3590 & -0.2406 \\
            -6.1207 &  0.3667 &  0.2545 &  0.0271 & -0.0218 &  0.0218 & -0.0271 & -0.2545 & -0.3667 \\
            -6.1207 &  0.3305 &  0.3141 &  0.0508 &  0.0666 &  0.0666 &  0.0508 &  0.3141 &  0.3305 \\
            \hline\hline
            -3.2635 & -0.0775 &  0.0932 &  0.1540 & -0.1558 & -0.1558 &  0.1540 &  0.0932 & -0.0775 \\
            -3.2635 &  0.1282 & -0.1534 & -0.0135 &  0.1621 & -0.1621 &  0.0135 &  0.1534 & -0.1282 \\
            -3.2635 &  0.1093 & -0.0221 & -0.2236 & -0.1260 &  0.1260 &  0.2236 &  0.0221 & -0.1093 \\
            -3.2635 &  0.1895 &  0.1160 & -0.1725 & -0.1955 & -0.1955 & -0.1725 &  0.1160 &  0.1895 \\
            -3.2635 & -0.2128 &  0.2686 & -0.1676 &  0.1010 &  0.1010 & -0.1676 &  0.2686 & -0.2128 \\
            -3.2635 & -0.0792 &  0.2671 & -0.2318 &  0.2961 & -0.2961 &  0.2318 & -0.2671 &  0.0792 \\
            -3.2635 & -0.3642 & -0.2700 & -0.1621 &  0.0192 & -0.0192 &  0.1621 &  0.2700 &  0.3642 \\
            -3.2635 &  0.2823 &  0.2711 &  0.2207 &  0.2398 &  0.2398 &  0.2207 &  0.2711 &  0.2823 \\
            \hline\hline
            -2.0283 & -0.0558 &  0.0241 &  0.0519 & -0.1476 & -0.1476 &  0.0519 &  0.0241 & -0.0558 \\
            -2.0283 &  0.0560 & -0.0014 & -0.0312 & -0.1570 &  0.1570 &  0.0312 &  0.0014 & -0.0560 \\
            -2.0283 &  0.0139 & -0.0367 &  0.1923 & -0.0329 &  0.0329 & -0.1923 &  0.0367 & -0.0139 \\
            -2.0283 &  0.0125 &  0.0262 &  0.1890 &  0.0661 &  0.0661 &  0.1890 &  0.0262 &  0.0125 \\
            -2.0283 & -0.1986 &  0.2569 & -0.0565 &  0.0972 &  0.0972 & -0.0565 &  0.2569 & -0.1986 \\
            -2.0283 & -0.1990 &  0.2713 &  0.0519 & -0.0836 &  0.0836 & -0.0519 & -0.2713 &  0.1990 \\
            -2.0283 & -0.3065 & -0.2471 & -0.0420 & -0.0987 &  0.0987 &  0.0420 &  0.2471 &  0.3065 \\
            -2.0283 &  0.3072 &  0.2627 & -0.0279 & -0.0829 & -0.0829 & -0.0279 &  0.2627 &  0.3072 \\
            \hline\hline
             2.0289 & -0.0558 & -0.0243 & -0.0525 & -0.1476 & -0.1476 & -0.0525 & -0.0243 & -0.0558 \\
             2.0289 &  0.0560 &  0.0012 &  0.0319 & -0.1572 &  0.1572 & -0.0319 & -0.0012 & -0.0560 \\
             2.0289 &  0.0142 &  0.0368 & -0.1924 & -0.0338 &  0.0338 &  0.1924 & -0.0368 & -0.0142 \\
             2.0289 &  0.0127 & -0.0261 & -0.1892 &  0.0668 &  0.0668 & -0.1892 & -0.0261 &  0.0127 \\
             2.0289 & -0.1988 & -0.2570 &  0.0565 &  0.0974 &  0.0974 &  0.0565 & -0.2570 & -0.1988 \\
             2.0289 & -0.1992 & -0.2714 & -0.0518 & -0.0837 &  0.0837 &  0.0518 &  0.2714 &  0.1992 \\
             2.0289 & -0.3066 &  0.2475 &  0.0420 & -0.0988 &  0.0988 & -0.0420 & -0.2475 &  0.3066 \\
             2.0289 &  0.3073 & -0.2630 &  0.0278 & -0.0829 & -0.0829 &  0.0278 & -0.2630 &  0.3073 \\
            \hline\hline
             3.2663 & -0.0775 & -0.0932 & -0.1540 & -0.1558 & -0.1558 & -0.1540 & -0.0932 & -0.0775 \\
             3.2663 &  0.1282 &  0.1533 &  0.0134 &  0.1622 & -0.1622 & -0.0134 & -0.1533 & -0.1282 \\
             3.2663 &  0.1093 &  0.0222 &  0.2237 & -0.1259 &  0.1259 & -0.2237 & -0.0222 & -0.1093 \\
             3.2663 &  0.1896 & -0.1160 &  0.1725 & -0.1954 & -0.1954 &  0.1725 & -0.1160 &  0.1896 \\
             3.2663 & -0.2128 & -0.2687 &  0.1675 &  0.1009 &  0.1009 &  0.1675 & -0.2687 & -0.2128 \\
             3.2663 & -0.0782 & -0.2680 &  0.2314 &  0.2960 & -0.2960 & -0.2314 &  0.2680 &  0.0782 \\
             3.2663 & -0.3645 &  0.2692 &  0.1628 &  0.0202 & -0.0202 & -0.1628 & -0.2692 &  0.3645 \\
             3.2663 &  0.2823 & -0.2710 & -0.2207 &  0.2399 &  0.2399 & -0.2207 & -0.2710 &  0.2823 \\
            \hline\hline
             6.1230 & -0.0038 & -0.0083 &  0.1688 & -0.1768 &  0.1768 & -0.1688 &  0.0083 &  0.0038 \\
             6.1230 & -0.0041 &  0.0055 &  0.1697 &  0.1761 &  0.1761 &  0.1697 &  0.0055 & -0.0041 \\
             6.1230 & -0.0532 & -0.0375 & -0.2099 & -0.1975 &  0.1975 &  0.2099 &  0.0375 &  0.0532 \\
             6.1230 & -0.0475 &  0.0267 & -0.2114 &  0.2018 &  0.2018 & -0.2114 &  0.0267 & -0.0475 \\
             6.1230 & -0.2897 & -0.3096 &  0.0150 & -0.0115 & -0.0115 &  0.0150 & -0.3096 & -0.2897 \\
             6.1230 &  0.2402 &  0.3590 & -0.0549 & -0.0744 &  0.0744 &  0.0549 & -0.3590 & -0.2402 \\
             6.1230 &  0.3667 & -0.2539 & -0.0270 & -0.0218 &  0.0218 &  0.0270 &  0.2539 & -0.3667 \\
             6.1230 &  0.3302 & -0.3139 & -0.0507 &  0.0663 &  0.0663 & -0.0507 & -0.3139 &  0.3302 \\
            \hline
        \end{tabular}
    \end{center}
    \caption{Bath parameters for a cluster DMFT calculation on the 2d square lattice Hubbard model at half-filling and $U\ / \ t = 8$ for a (2$\times$4) cluster with $N_p = 6$. These were obtained with a linear frequency grid of 50 points in the interval $i\omega_n \in [0, 40\ t]$.}
    \label{Table:2x4_Np6}
\end{table}

\begin{table}[h!]
    \begin{center}
        \begin{tabular}{|r||c|c|c|c|c|c|}
        \hline
          \textbf{$\epsilon_\ell$} & \textbf{$V_{\ell,1}$} & \textbf{$V_{\ell,2}$} & \textbf{$V_{\ell,3}$} & \textbf{$V_{\ell,4}$} & \textbf{$V_{\ell,5}$} & \textbf{$V_{\ell,6}$} \\
          \hline\hline
            -3.0845 & -0.0911 &  0.2437 & -0.3632 &  0.3632 & -0.2437 &  0.0911 \\
            -3.0845 &  0.2508 & -0.4635 &  0.1729 &  0.1729 & -0.4635 &  0.2508 \\
            -3.0845 &  0.3327 & -0.4287 & -0.3711 &  0.3711 &  0.4287 & -0.3327 \\
            -3.0845 &  0.1917 & -0.1841 & -0.7716 & -0.7716 & -0.1841 &  0.1917 \\
            -3.0845 &  0.7695 &  0.4479 &  0.0843 &  0.0843 &  0.4479 &  0.7695 \\
            -3.0845 &  0.7524 &  0.4727 &  0.1284 & -0.1284 & -0.4727 & -0.7524 \\
            \hline\hline
             3.0845 & -0.0911 & -0.2437 & -0.3632 & -0.3632 & -0.2437 & -0.0911 \\
             3.0845 &  0.2508 &  0.4635 &  0.1729 & -0.1729 & -0.4635 & -0.2508 \\
             3.0845 &  0.3327 &  0.4287 & -0.3711 & -0.3711 &  0.4287 &  0.3327 \\
             3.0845 &  0.1917 &  0.1841 & -0.7716 &  0.7716 & -0.1841 & -0.1917 \\
             3.0845 & -0.7695 &  0.4479 & -0.0843 &  0.0843 & -0.4479 &  0.7695 \\
             3.0845 &  0.7524 & -0.4727 &  0.1284 &  0.1284 & -0.4727 &  0.7524 \\
            \hline
        \end{tabular}
    \end{center}
    \caption{Bath parameters for a cluster DMFT calculation on the 2d square lattice Hubbard model at half-filling and $U\ / \ t = 8$ for a (1$\times$6) cluster with $N_p = 2$. These were obtained with a linear frequency grid of 50 points in the interval $i\omega_n \in [0, 40\ t]$.}
    \label{Table:1x6_Np2}
\end{table}

\begin{table}[h!]
    \begin{center}
        \begin{tabular}{|r||c|c|c|c|c|c|}
        \hline
          \textbf{$\epsilon_\ell$} & \textbf{$V_{\ell,1}$} & \textbf{$V_{\ell,2}$} & \textbf{$V_{\ell,3}$} & \textbf{$V_{\ell,4}$} & \textbf{$V_{\ell,5}$} & \textbf{$V_{\ell,6}$} \\
          \hline\hline
            -3.1598 &  0.0920 & -0.2151 &  0.3126 & -0.3126 &  0.2151 & -0.0920 \\
            -3.1598 & -0.2544 &  0.3830 & -0.1492 & -0.1492 &  0.3830 & -0.2544 \\
            -3.1598 & -0.3702 &  0.2711 &  0.2955 & -0.2955 & -0.2711 &  0.3702 \\
            -3.1598 &  0.4890 &  0.1571 & -0.4304 & -0.4304 &  0.1571 &  0.4890 \\
            -3.1598 & -0.5331 & -0.5138 & -0.1965 &  0.1965 &  0.5138 &  0.5331 \\
            -3.1598 &  0.3478 &  0.4488 &  0.5588 &  0.5588 &  0.4488 &  0.3478 \\
            \hline\hline
            -1.4543 &  0.0164 & -0.0414 &  0.0530 & -0.0530 &  0.0414 & -0.0164 \\
            -1.4543 & -0.0230 &  0.1180 &  0.0440 &  0.0440 &  0.1180 & -0.0230 \\
            -1.4543 & -0.1007 &  0.0417 & -0.1646 & -0.1646 &  0.0417 & -0.1007 \\
            -1.4543 &  0.0359 & -0.1681 & -0.1424 &  0.1424 &  0.1681 & -0.0359 \\
            -1.4543 & -0.3515 & -0.1007 &  0.0302 & -0.0302 &  0.1007 &  0.3515 \\
            -1.4543 &  0.3498 &  0.1353 & -0.1798 & -0.1798 &  0.1353 &  0.3498 \\
            \hline\hline
             1.4545 &  0.0164 &  0.0414 &  0.0530 &  0.0530 &  0.0414 &  0.0164 \\
             1.4545 & -0.0230 & -0.1180 &  0.0440 & -0.0440 &  0.1180 &  0.0230 \\
             1.4545 & -0.1007 & -0.0417 & -0.1646 &  0.1646 &  0.0417 &  0.1007 \\
             1.4545 & -0.0359 & -0.1681 &  0.1424 &  0.1424 & -0.1681 & -0.0359 \\
             1.4545 &  0.3516 & -0.1007 & -0.0302 & -0.0302 & -0.1007 &  0.3516 \\
             1.4545 & -0.3498 &  0.1353 &  0.1798 & -0.1798 & -0.1353 &  0.3498 \\
            \hline\hline
             3.1599 & -0.0920 & -0.2151 & -0.3126 & -0.3126 & -0.2151 & -0.0920 \\
             3.1599 & -0.2544 & -0.3830 & -0.1492 &  0.1492 &  0.3830 &  0.2544 \\
             3.1599 &  0.3702 &  0.2710 & -0.2955 & -0.2955 &  0.2710 &  0.3702 \\
             3.1599 & -0.4890 &  0.1571 &  0.4304 & -0.4304 & -0.1571 &  0.4890 \\
             3.1599 &  0.5331 & -0.5138 &  0.1965 &  0.1965 & -0.5138 &  0.5331 \\
             3.1599 & -0.3478 &  0.4488 & -0.5589 &  0.5589 & -0.4488 &  0.3478 \\
            \hline
        \end{tabular}
    \end{center}
    \caption{Bath parameters for a cluster DMFT calculation on the 2d square lattice Hubbard model at half-filling and $U\ / \ t = 8$ for a (1$\times$6) cluster with $N_p = 4$. These were obtained with a Matsubara frequency grid with $\beta = 600$ and cut-off frequency $i\omega_c = 2\ t$.}
    \label{Table:1x6_Np4}
\end{table}

\begin{table}[h!]
    \begin{center}
        \begin{tabular}{|r||c|c|c|c|c|c|}
        \hline
          \textbf{$\epsilon_\ell$} & \textbf{$V_{\ell,1}$} & \textbf{$V_{\ell,2}$} & \textbf{$V_{\ell,3}$} & \textbf{$V_{\ell,4}$} & \textbf{$V_{\ell,5}$} & \textbf{$V_{\ell,6}$} \\
          \hline\hline
            -6.0606 & -0.0106 & -0.2544 &  0.2430 &  0.2430 & -0.2544 & -0.0106 \\
            -6.0606 &  0.0001 &  0.2580 & -0.2824 &  0.2824 & -0.2580 & -0.0001 \\
            -6.0606 & -0.0279 &  0.3157 &  0.2884 & -0.2884 & -0.3157 &  0.0279 \\
            -6.0606 & -0.1236 &  0.3260 &  0.3359 &  0.3359 &  0.3260 & -0.1236 \\
            -6.0606 &  0.5724 &  0.0275 &  0.0254 & -0.0254 & -0.0275 & -0.5724 \\
            -6.0606 &  0.5607 &  0.0902 &  0.1189 &  0.1189 &  0.0902 &  0.5607 \\
            \hline\hline
            -3.0346 &  0.0862 & -0.1539 &  0.2256 & -0.2256 &  0.1539 & -0.0862 \\
            -3.0346 & -0.2267 &  0.2654 & -0.0821 & -0.0821 &  0.2654 & -0.2267 \\
            -3.0346 & -0.2831 &  0.1814 &  0.2319 & -0.2319 & -0.1814 &  0.2831 \\
            -3.0346 &  0.3887 &  0.2201 & -0.3617 & -0.3617 &  0.2201 &  0.3887 \\
            -3.0346 & -0.4388 & -0.4802 & -0.1601 &  0.1601 &  0.4802 &  0.4388 \\
            -3.0346 &  0.2653 &  0.3880 &  0.5212 &  0.5212 &  0.3880 &  0.2653 \\
            \hline\hline
            -1.5552 & -0.0236 &  0.0639 & -0.0766 &  0.0766 & -0.0639 &  0.0236 \\
            -1.5552 &  0.0161 & -0.1401 & -0.0702 & -0.0702 & -0.1401 &  0.0161 \\
            -1.5552 & -0.1187 &  0.0726 & -0.1720 & -0.1720 &  0.0726 & -0.1187 \\
            -1.5552 & -0.0358 &  0.1907 &  0.1703 & -0.1703 & -0.1907 &  0.0358 \\
            -1.5552 & -0.3716 & -0.0987 &  0.0324 & -0.0324 &  0.0987 &  0.3716 \\
            -1.5552 &  0.3674 &  0.1398 & -0.1945 & -0.1945 &  0.1398 &  0.3674 \\
            \hline\hline
             1.5555 &  0.0237 &  0.0640 &  0.0766 &  0.0766 &  0.0640 &  0.0237 \\
             1.5555 &  0.0161 &  0.1401 & -0.0702 &  0.0702 & -0.1401 & -0.0161 \\
             1.5555 &  0.1187 &  0.0726 &  0.1720 & -0.1720 & -0.0726 & -0.1187 \\
             1.5555 &  0.0359 &  0.1907 & -0.1703 & -0.1703 &  0.1907 &  0.0359 \\
             1.5555 &  0.3716 & -0.0987 & -0.0323 & -0.0323 & -0.0987 &  0.3716 \\
             1.5555 & -0.3674 &  0.1398 &  0.1945 & -0.1945 & -0.1398 &  0.3674 \\
            \hline\hline
             3.0348 & -0.0862 & -0.1539 & -0.2256 & -0.2256 & -0.1539 & -0.0862 \\
             3.0348 & -0.2267 & -0.2654 & -0.0821 &  0.0821 &  0.2654 &  0.2267 \\
             3.0348 &  0.2831 &  0.1814 & -0.2319 & -0.2319 &  0.1814 &  0.2831 \\
             3.0348 & -0.3887 &  0.2201 &  0.3617 & -0.3617 & -0.2201 &  0.3887 \\
             3.0348 &  0.4387 & -0.4802 &  0.1601 &  0.1601 & -0.4802 &  0.4387 \\
             3.0348 & -0.2653 &  0.3879 & -0.5212 &  0.5212 & -0.3879 &  0.2653 \\
            \hline\hline
             6.0607 & -0.0106 &  0.2544 &  0.2431 & -0.2431 & -0.2544 &  0.0106 \\
             6.0607 &  0.0001 & -0.2580 & -0.2824 & -0.2824 & -0.2580 &  0.0001 \\
             6.0607 & -0.0279 & -0.3157 &  0.2884 &  0.2884 & -0.3157 & -0.0279 \\
             6.0607 & -0.1236 & -0.3260 &  0.3358 & -0.3358 &  0.3260 &  0.1236 \\
             6.0607 &  0.5724 & -0.0275 &  0.0254 &  0.0254 & -0.0275 &  0.5724 \\
             6.0607 & -0.5607 &  0.0902 & -0.1188 &  0.1188 & -0.0902 &  0.5607 \\
            \hline
        \end{tabular}
    \end{center}
    \caption{Bath parameters for a cluster DMFT calculation on the 2d square lattice Hubbard model at half-filling and $U\ / \ t = 8$ for a (1$\times$6) cluster with $N_p = 6$. These were obtained with a linear frequency grid of 50 points in the interval $i\omega_n \in [0, 40\ t]$.}
    \label{Table:1x6_Np6}
\end{table}

\FloatBarrier

\subsection*{Comparison to previous studies}

\begin{figure}[b]
\includegraphics{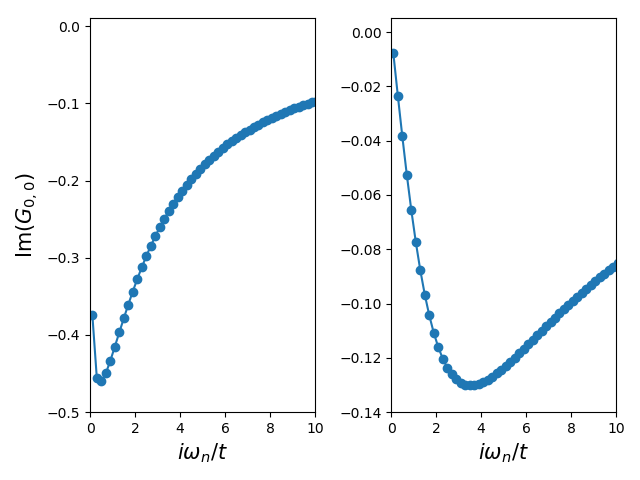}
\caption{Imaginary part of the first diagonal element of the Green's function along the imaginary frequency axis for the one-dimensional Hubbard model at half filling with $U/t = 2$ (left) and $U/t = 8$ (right). These correspond to cluster DMFT calculations with $N_c = 4$ and $N_b = 8$, i.e. 4 poles. They are in great agreement with Fig.~5 in~\cite{Go2017}. The fit was computed in a linear frequency grid with 200 points between $i\omega_{min} = 0.1$ $t$ and $i\omega_{max} = 40$ $t$.}
\label{fig:1dImag}
\end{figure}

\begin{figure}[b]
\includegraphics{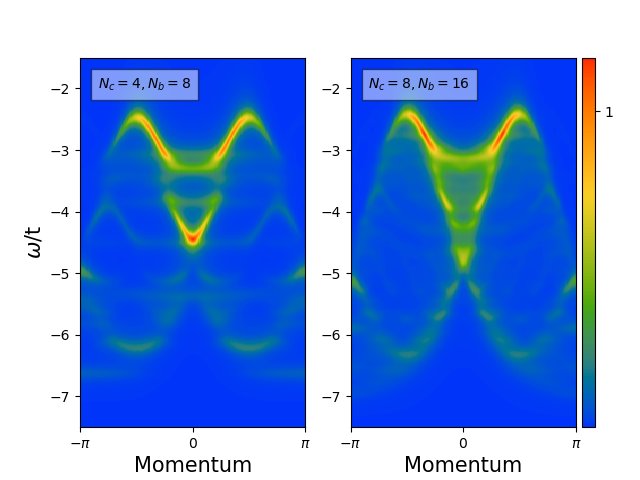}
\caption{Spectral weights $A(k,\omega)$ for the one-dimensional Hubbard model at half filling with $U/t = 8$. These correspond to cluster DMFT calculations with $N_c = 4$ and $N_b = 8$, i.e. 4 poles, (left) and $N_c = 8$ and $N_b = 16$, i.e. 8 poles, (right). They are in great agreement with Fig.~3 in~\cite{Go2017}. The fit was computed in a linear frequency grid with 200 points between $i\omega_{min} = 0.1$ $t$ and $i\omega_{max} = 40$ $t$.}
\label{fig:1dReal}
\end{figure}

\begin{figure}[b]
\includegraphics{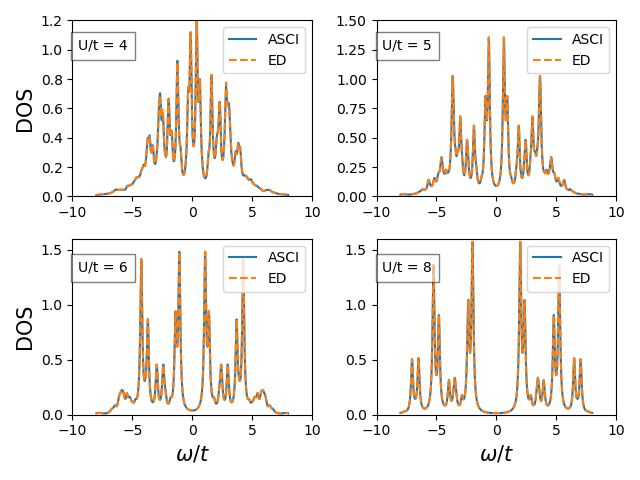}
\caption{Linear densities of states (i.e. $-\frac{1}{\pi}\mathrm{tr(Im(}G(\omega)))$) for the two-dimensional Hubbard model at half filling with several $U/t$ values. These correspond to cluster DMFT calculations with $N_c = 2\times2$ and $N_b = 8$, i.e. 2 poles. The blue solid line was computed using ASCI as impurity solver, the orange dashed line using ED. They are in good agreement with Fig.~8 in~\cite{Zgid2012}. The fit was computed in a linear frequency grid with 1000 points between $i\omega_{min} = 0.02$ $t$ and $i\omega_{max} = 40$ $t$.}
\label{fig:2dReal}
\end{figure}

\begin{figure}[b]
\includegraphics{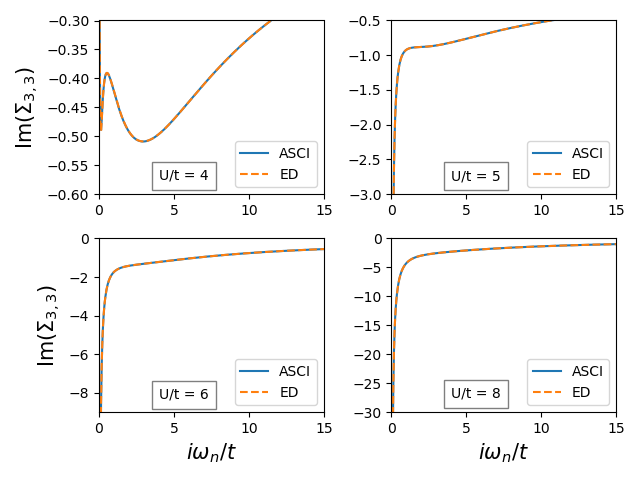}
\caption{Imaginary part of a diagonal component of the self energy along the imaginary frequency axis for the two-dimensional Hubbard model at half filling with several $U/t$ values. These results correspond to cluster DMFT calculations with $N_c = 2\times2$ and $N_b = 8$, i.e. 2 poles. The blue solid line was computed using ASCI as impurity solver, the orange dashed line using ED. They are in good agreement with Fig.~9 in~\cite{Zgid2012}. The fit was computed in a linear frequency grid with 1000 points between $i\omega_{min} = 0.02$ $t$ and $i\omega_{max} = 40$ $t$.}
\label{fig:2dImag}
\end{figure}

Here we provide comparison of the results that can be obtained with our method to previous zero temperature ED based impurity solvers, in particular the configuration interaction based solvers in Ref.~\cite{Zgid2012,Go2017}. In Fig.~\ref{fig:1dImag} and~\ref{fig:1dReal} we show Green's functions along the imaginary frequency axis and spectral weights $A(k,\omega)$ for one-dimensional Hubbard model calculations at half filling, which show excellent comparison with~\cite{Go2017}. These are calculations with $N_c = 4, 8$ and $N_b = 8, 16$ respectively. Since the number of boundary sites in the cluster is $N_c' = 2$ for both cases, the $N_b = 8$ corresponds to a fit with $N_p = 4$ poles, while the $N_b = 16$ corresponds to a fit with $N_p = 8$ poles. As comparisons for the two-dimensional Hubbard model at half filling, we provide results for the linear density of states and the imaginary part of the diagonal self energy $\Sigma$ along the imaginary frequency axis for several interaction values $U / t$ in Fig.~\ref{fig:2dReal} and~\ref{fig:2dImag}. We provide results using our impurity solver, ASCI, as blue solid lines and using an ED solver as orange dashed lines. In order to make ED feasible, we limit ourselves to $N_c = 2\times 2$ and $N_b = 8$, i.e. a fit with $N_p = 2$ poles. We observe excellent agreement between both, which in turn are in good agreement with the results in~\cite{Zgid2012}. All the calculations using ASCI as impurity solver employed a target space of $10^5$ determinants.

\end{document}